\pdfoutput=1
\documentclass[aps,prb,showpacs,twocolumn,longbibliography]{revtex4}
\usepackage{graphicx}
\usepackage{amssymb}
\usepackage{amsmath}
\usepackage{color}
\usepackage{dsfont}
\usepackage{paralist}
\usepackage{bm}
\usepackage{hyperref}
\usepackage{verbatim}
\usepackage{textcomp}

\def\cL{{\cal L}}
\def\cO{{\cal O}}

\def\bk{{\bf k}}

\def\bq{{\bf q}}
\def\bp{{\bf p}}
\def\br{{\bf r}}

\def\hbsigma{\hat{\boldsymbol \sigma}}

\def\bnabla{{\boldsymbol \nabla}}

\def\holOne{\mathds{1}}

\begin{document}

\title{Critical exponents at the unconventional disorder-driven transition in a Weyl semimetal}


\author{S.V.~Syzranov$^{1,2}$, P.M.~Ostrovsky$^{3,4}$, V.~Gurarie$^{1,2}$, L.~Radzihovsky$^{1,2,5}$}

\affiliation{
$^1$Physics Department, University of Colorado, Boulder, Colorado 80309, USA
\\
$^2$Center for Theory of Quantum Matter, University of Colorado, Boulder, Colorado 80309, USA
\\
$^3$Max Planck Institute for Solid State Research, Heisenbergstr. 1, 70569 Stuttgart, Germany
\\
$^4$L.D.~Landau Institute for Theoretical Physics RAS, 119334 Moscow, Russia
\\
$^5$JILA, NIST, University of Colorado, Boulder, Colorado 80309, USA
}

\begin{abstract}
  {
  Disordered non-interacting systems in sufficiently high dimensions
  have been predicted to display a non-Anderson disorder-driven transition that
  manifests itself in the critical behaviour of the density of states and other physical observables.
  Recently the critical properties of this transition have been extensively studied for the specific case
  of Weyl semimetals by means of numerical and renormalisation-group approaches. 
  Despite this,
  the values of the critical exponents at such a transition in a Weyl semimetal are currently under debate.  
  We present an independent calculation of the critical exponents
  using a two-loop renormalisation-group approach for Weyl fermions in $2-\varepsilon$ dimensions
  and resolve controversies currently existing in the literature.
  }
\end{abstract}



\date{\today}
\maketitle

It was proposed\cite{Fradkin1,Fradkin2} 30 years ago that three-dimensional (3D)
disordered systems with Weyl and Dirac quasiparticle dispersion can display an unconventional disorder-driven
transition that lies in a non-Anderson universality class. In particular, in contrast with the 
Anderson localisation transition,
the density of states at this transition has been suggested\cite{Fradkin1} to
display a critical behaviour, with the scaling function proposed in Ref.~\onlinecite{Herbut}.

Recently we have demonstrated\cite{Syzranov:Weyl,Syzranov:unconv} that such transitions
occur near nodes and band edges in {\it all} materials in sufficiently
{\it high dimensions} $d$ and are not unique to Dirac (Weyl) systems. 
In systems that allow for localisation the transition manifests itself also in the
unusual behaviour of the mobility threshold\cite{Syzranov:unconv}.
As the concept of high dimensions here is defined relative to the quasiparticle dispersion,
possible playgrounds include a number of systems in physical $d=1,2,3$ dimensions, with higher
dimensions being accessible numerically. For example, recently we have shown how this
transition can be observed in 1D and 2D arrays of ultracold ions in optical
or magnetic traps\cite{Garttner:longrange}.

Because Weyl semimetals (WSMs) are currently one of the most well-known and experimentally accessible 
platforms\cite{ZHasan:TaAs,ZHasan:TaAs2,Weng:PhotCrystWSM} for the observation of these high-dimensional disorder-driven phenomena, tremendous research
efforts have been directed at studying the critical properties of the transition in a WSM.
Nevertheless, the values of the critical exponents at the transition are currently under debate. 

Quantum criticality near the transition has been studied analytically for Dirac and Weyl particles
in dimensions $d>2$ by means of perturbative renormalisation-group
approaches\textsuperscript{\onlinecite{Goswami:TIRG},\onlinecite{Syzranov:Weyl,Syzranov:unconv},
\onlinecite{RoyDasSarma}},
large-$N$ (large number of valleys or particle flavours) analysis\cite{RyuNomura},
and (uncontrolled) self-consistent Born approximation\cite{Fradkin1,ShindouMurakami,OminatoKoshino} (SCBA),
equivalent to a large-$N$ calculation
in the limit $N=\infty$.

Usually Weyl and Dirac materials have only several nodes, $N\sim {\cal O}(1)$,
which makes large-$N$ approaches quantitatively inaccurate in all
dimensions $d$ [for detailed criticism of the SCBA vs. RG see Refs.~\onlinecite{AleinerEfetov,OstrovskyGornyMirlin},
\onlinecite{Syzranov:Weyl} (although Refs.~\onlinecite{AleinerEfetov} and \onlinecite{OstrovskyGornyMirlin}
are devoted to graphene, their arguments apply as well to Dirac and Weyl particles in all dimensions)].

Perturbative renormalisation-group (RG) analysis are controlled by the small parameter $\varepsilon=2-d$. 
Although 
such RG analysis for small $\varepsilon$ is not guaranteed to be quantitatively accurate when analytically
continued to 3D ($\varepsilon = -1$), already the
one-loop results\cite{Goswami:TIRG}\cite{Syzranov:Weyl,Syzranov:unconv,RoyDasSarma} predict the
correlation-length exponent $\nu=1$ and the dynamical exponent $z=3/2$  that
lie within $15\%$ and several percent of the values obtained numerically in
Refs.~\onlinecite{Herbut}, \onlinecite{LiuOhtsuki:LateNumerics}, \onlinecite{Bera:Weyl}.

Also, numerical analysis of Ref.~\onlinecite{Garttner:longrange}, albeit carried out
for 1D chiral systems, suggest that one-loop RG results accurately describe the critical properties
of such type of transitions.
Recently a value of $z = 1.49\pm 0.02$ very close to the one-loop result $z=3/2$ 
has been obtained
numerically in Ref.~\onlinecite{Brouwer:exponents}.
However, the same simulations found a value
of $\nu = 1.47\pm 0.03$ very different from the one-loop prediction $\nu=1$.
It has been argued in Ref.~\onlinecite{Pixley:ExactZ} that the one-loop result $z=3/2$
for the dynamical critical exponent is exact and holds in all orders of the renormalisation,
in contradiction with the analytical two-loop calculations
of Ref.~\onlinecite{RoyDasSarma} 
predicting\footnote{We note that in Ref.~\onlinecite{RoyDasSarma}
the parameter $\varepsilon$ is defined as $\varepsilon=d-2$, while our convention in this paper
is $\varepsilon=2-d$.}
$\nu=\left[-\varepsilon-\frac{\varepsilon^2}{8}+\ldots\right]^{-1}\approx 1.14$
and $z=1-\frac{\varepsilon}{2}-\frac{3}{16}\varepsilon^2+\ldots\approx 1.31$.
Similar analytical two-loop RG calculations for graphene\cite{Ostrovsky:grapheneRG} and 
for related the Gross-Neveu model\cite{Wetzel:twoloop,Ludwig:twoloop,Rossi1,Rossi2,TracasVlachos,Ludwig:Thirring}, with the results 
for the latter being well established in the high-energy literature,
yield beta-functions inconsistent with those of Refs.~\onlinecite{RoyDasSarma}
and \onlinecite{Pixley:ExactZ}.

In this paper we present an independent calculation of the critical exponents for the transition in a Weyl semimetal
using a two-loop renormalisation group approach for Weyl fermions in $2-\varepsilon$ dimensions
and resolve controversies currently existing in the literature. We obtain beta functions consistent with those
obtained for graphene in Ref.~\onlinecite{Ostrovsky:grapheneRG}
and for Gross-Neveu model in Refs.~\onlinecite{Wetzel:twoloop,Ludwig:twoloop,Rossi1,Rossi2,TracasVlachos,Ludwig:Thirring},
but in disagreement with the results of Refs.~\onlinecite{RoyDasSarma} and \onlinecite{Pixley:ExactZ}.
We find the critical exponents to be
\begin{align}
	\nu &=
	\left[-\varepsilon+\frac{\varepsilon^2}{2}+\ldots
	\right]^{-1},
	\label{Nu}
	\\
	z &=1-\frac{\varepsilon}{2}-\frac{\varepsilon^2}{8}+\ldots.
	\label{Z}
\end{align}

For a 3D Weyl semimetal ($\varepsilon=-1$) Eqs.~(\ref{Nu}) and (\ref{Z}) give 
\begin{align}
	\nu\approx 0.67, \quad z\approx 1.4.
\end{align}
We emphasise, however, that higher-loop corrections for $\varepsilon=-1$ will lead to deviations from these values
obtained from an RG calculation controlled by small $\varepsilon$.

The critical density of states in a WSM with short-range-correlated disorder
can be described in the supersymmetric representation\cite{Efetov:book}
by a field theory with the action
\begin{align}
	\cL=-i\int \Phi^\dagger	\left[i\omega-\hbsigma\hat\bk\right]\Phi\: d\br
	+\frac{1}{2}\varkappa_0\int(\Phi^\dagger\Phi)^2d\br,
	\label{Action0}
\end{align}
where $\Phi=(\chi\: s)^T$ is a vector consisting of an anticommuting (fermionic)
$\chi$ and commuting $s$ (bosonic) components, $\omega>0$ is the Matsubara frequency in the upper half-plane,
$\hbsigma$ is a vector of matrices generating a $d$-dimensional Clifford
algebra ($\{\hat\sigma_\alpha,\hat\sigma_\beta\}=2\delta_{\alpha\beta}\holOne$),
$\bk=-i\bnabla$ is the momentum operator,
and $\varkappa_0=\int\left<U(\br)U(\br^\prime)\right>_{\text{dis}}d\br^\prime$ is the strength of the short-range-correlated random disorder potential under consideration.
Equivalent field theories can be
derived also in Keldysh\cite{Kamenev:book} and replica\cite{BelitzKirkpatrick} representations.

We emphasise, that we express the action (\ref{Action0}) in terms of a positive Matsubara frequency $\omega$,
in contrast with the conventional real-frequency representation\cite{Efetov:book},
in order 
to regularise integrals in the renormalisation scheme used below.
Also, frequency $\omega$ in Eq.~\eqref{Action0}
plays the same role as the mass $m$ in the related Gross-Neveu model\cite{Ludwig:twoloop,Rossi1,Rossi2}.
The density of states is determined by the retarded Green's 
function $G^R(E,\br,\br^\prime)$ that can be obtained from the action (\ref{Action0})
using analytic continuation to real frequencies
$i\omega\rightarrow E+i0$.

We note, that realistic materials always have an even number of Weyl nodes, 
due to the fermion doubling theorem\cite{NielsenNinomiya}, and thus in general should be described by an action  
with an even number of Weyl fermion flavours. However, for sufficiently smooth disorder
internodal scattering can be neglected, and the material is equivalent to an even number of copies
of single-node WSMs described by the action (\ref{Action0}).

In dimensions $d>2$ this field theory leads to ultraviolet divergencies in physical observables
and requires an appropriate RG treatment.

We study the behaviour of the system at frequency $\omega$ and at long length scales, $\bk\rightarrow0$,
following the {\it minimal subtraction} renormalisation scheme\cite{PeskinSchroeder}. We use dimensional regularisation
by computing observables in lower $d=2-\varepsilon$ dimensions (with small $\varepsilon>0$) and then analytically
continue renormalised observables to the higher dimensions of interest ($\varepsilon<0$).  
The Lagrangian~\eqref{Action0} in this scheme is separated into the effective Lagrangian $\cL_E$
of variables observable in the long-wave limit of interest and the counterterms:
\begin{align}
	\cL & =\cL_E+\cL_{\text{counter}},
	\label{Action1}\\
	\cL_E &=-i\int \psi^\dagger	\left(i\Omega-\hbsigma\hat\bk\right)\psi\: d\br
	+\frac{1}{2}\varkappa\int(\psi^\dagger\psi)^2d\br,
	\label{LE}
\end{align}
where the energy scale $\Omega$ and the renormalised disorder strength $\varkappa$ are experimentally
observable, and the counterterms
$\cL_{\text{counter}}$ cancel the divergent (in the powers of $1/\varepsilon$) contributions
to physical observables that come from the Lagrangian~\eqref{LE}. 


The strength of disorder can be conveniently characterised by 
the dimensionless parameter\cite{Syzranov:unconv}
\begin{align}
	\gamma=2C_d\varkappa\,\Omega^{-\varepsilon},
	\label{DimensDisorder}
\end{align}
where $C_d=2^{1-d}\pi^{-\frac{d}{2}}/\Gamma\left(\frac{d}{2}\right)$. For a given ``bare'' disorder strength $\varkappa_0$
the renormalised dimensionless disorder strength $\gamma$ and the characteristic energy $\Omega$ of the long-wave
behaviour of disorder-averaged observables are related by the RG equation (see Appendix for a detailed derivation)
\begin{align}
	\frac{\partial\gamma}{\partial\ln\Omega}
	=-\varepsilon\gamma-\gamma^2-\frac{1}{2}\gamma^3+\ldots.
	\label{GammaRGE}
\end{align}
The dependence of $\Omega$ on the frequency $\omega$
is described by the RG equation
\begin{align}
	\left(\frac{\partial\ln\Omega}{\partial\ln\omega}\right)^{-1}
	=1+\frac{\gamma}{2}+\frac{\gamma^2}{8}+\ldots.
	\label{OmegaRGE}
\end{align}

Our two-loop RG equations \eqref{GammaRGE} and \eqref{OmegaRGE} are consistent with the previous studies 
of Weyl fermions in $d=2-\varepsilon$ dimensions: in the framework of
Gross-Neveu model in Refs.~\onlinecite{Wetzel:twoloop,Ludwig:twoloop,Rossi1,Rossi2,TracasVlachos,Ludwig:Thirring}
and of graphene
in Ref.~\onlinecite{Ostrovsky:grapheneRG}.

Eq.~\eqref{GammaRGE} shows that the dimensionless disorder strength grows or decreases depending on
whether or not it exceeds a critical disorder strength
\begin{align}
	\gamma_c=-\varepsilon-\frac{\varepsilon^2}{2}+\ldots.
\end{align}
The existence of such a repulsive fixed point signals a transition, discussed
in the 
literature\cite{Fradkin1,Fradkin2,Goswami:TIRG,Syzranov:Weyl,Syzranov:unconv,Herbut,LiuOhtsuki:LateNumerics,Brouwer:exponents,Brouwer:WSMcond,RoyDasSarma,Pixley:ExactZ,Shapourian:PhaseDiagr},
between a weak-disorder
phase and a strong-disorder phase.

Using that near the transition $\frac{\partial\ln(\gamma-\gamma_c)}{\partial\ln\Omega}=-\nu^{-1}$ and
Eq.~\eqref{GammaRGE}
we obtain the correlation-length critical exponent \eqref{Nu}.
The divergence of $\nu$ in the $\varepsilon\rightarrow0$ limit reflects the fact that $d=2$ ($\varepsilon=0$)
is the lower critical dimension for this transition.

Our analysis predicts $\omega\propto\Omega^z$ at the critical disorder strength $\gamma=\gamma_c$.
Eq.~\eqref{OmegaRGE} then gives the dynamical critical exponent \eqref{Z}.

We note that for small $\varepsilon\ll1$ the correlation-length exponent $\nu$, Eq.~\eqref{Nu}, satisfies
Harris criterion (Chayes inequality)\cite{ChayesChayes,Harris} $\nu\geq2/d$. 
Because the correlation length
$\xi\propto\Omega^{-1}\propto\omega^{-\frac{1}{z}}$ can be measured
as a function
of the energy $\omega$ for $\varkappa=\varkappa_c$ as well as a function of disorder strength $\xi\propto|\varkappa-\varkappa_c|^{-\nu}$
for $\omega=0$, a similar criterion can be applied heuristically to the
exponent $\tilde{\nu}=1/z$, yielding $z\leq d/2$, consistent with
our result Eq.~\eqref{Z}.


In conclusion, we have presented a two-loop renormalisation-group 
analysis of the critical properties of the 
unconventional disorder-driven transition for Weyl fermions above two dimensions
and found the correlation-length and dynamical critical exponents, Eqs.~\eqref{Nu} and \eqref{Z}.
The beta-functions 
that we obtain are consistent with those for graphene and Gross-Neveu models
studied previously in the 
literature\footnote{
We note, that the sign of the coupling $\gamma$ in the disordered problem under consideration
is opposite to that of the respective models studied in high-energy literature for repulsively interacting fermions, which
reflects in different signs of even-in-$\gamma$ terms of beta-functions.
}.

{\it Acknowledgements.}
We appreciate useful discussions with A.W.W.~Ludwig.
Our work was supported by the Alexander von Humboldt
Foundation through the Feodor Lynen Research Fellowship (SVS) and by the NSF grants
DMR-1001240 (LR and SVS),
DMR-1205303(VG and SVS), PHY-1211914 (VG and SVS), and PHY-1125844 (SVS).
LR also acknowledges support by the Simons Investigator award from the Simons Foundation.

{\it Note added.}
After posting this paper on arXiv, the results of Refs.~\onlinecite{Pixley:ExactZ} and
\onlinecite{RoyDasSarma}, contradicting our conclusions here, have been withdrawn by their authors;
the claim of $z=3/2$ being the exact dynamical exponent has been removed\cite{Pixley:ExactZpubliahed}
by the authors of Ref.~\onlinecite{Pixley:ExactZ}, 
while the results of Ref.~\onlinecite{RoyDasSarma} for the critical exponents
have been retracted
in erratum~\onlinecite{RoyDasSarma:erratum}.



\onecolumngrid
\appendix

\section{Renormalisation scheme}

Disorder-averaged observables, e.g., the density of states or conductivity, calculated perturbatively in disorder
strength using action \eqref{Action0} in dimensions $d>2$ contain ultravioletly-divergent
contributions
that require an appropriate
renormalisation-group treatment.

In this paper we use the minimal-subtraction renormalisation-group scheme\cite{PeskinSchroeder}.
The respective integrals in this scheme are evaluated in lower $d=2-\varepsilon$
dimensions ($\varepsilon>0$), to ensure their ultraviolet convergence,
making analytic continuation to higher dimensions ($\varepsilon<0$)
in the end of the calculation. Also, as we show below, the infrared convergence of momentum
integrals is ensured by using Matsubara frequencies $i\Omega$ in place of real frequencies. 
The renormalisation procedure consists in calculating perturbative corrections to the
disorder-free particle propagator 
\begin{align}
	G(i\Omega,\bp)=(i\Omega-\hbsigma\bp)^{-1}=-\frac{i\Omega+\hbsigma\bp}{\Omega^2+p^2}
\end{align}
and the coupling $\varkappa$ in the Lagrangian \eqref{LE} and adding counterterms $\cL_{\text{counter}}$
to the Lagrangian
in order to cancel divergent (in powers of $1/\varepsilon$) contributions.
The renormalised quantities $\varkappa$ and $\Omega$ can then be related to the ``bare''
$\varkappa_0$ and $\omega$ by comparing the initial Lagrangian~\eqref{Action0} and the Lagrangian~\eqref{Action1}
expressed in the renormalised variables.

\begin{figure}[htbp]
	\centering
	\includegraphics[width=0.37\textwidth]{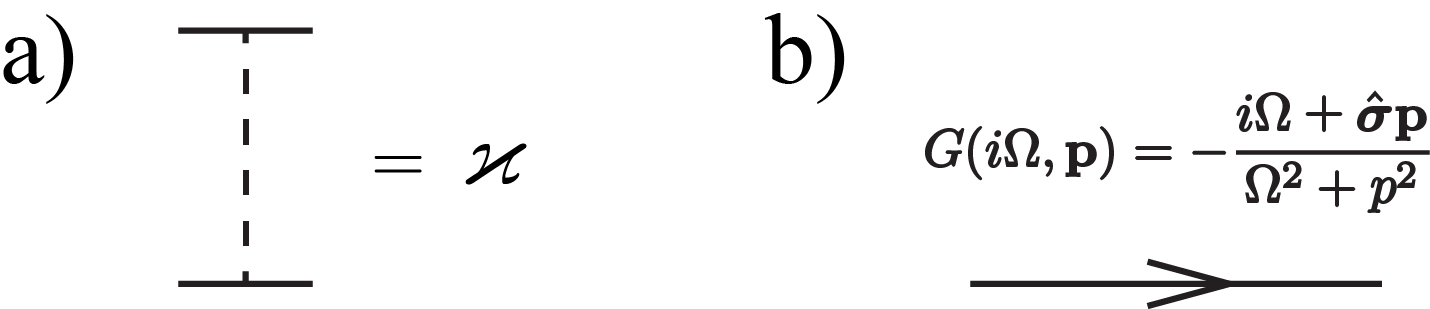}
	\caption{Elements of the diagrammatic technique: a) impurity line and b) propagator.}
	\label{DiagrElem}
\end{figure}
Perturbative corrections to the propagator and disorder strength can be obtained straightforwardly
using the Lagrangian~\eqref{LE}. For convenience we utilise the conventional disorder-averaging
diagrammatic technique\cite{AGD}, Fig.~\ref{DiagrElem}. The impurity line, Fig~\ref{DiagrElem}a,
is a tensor product of two operators $\hat\tau_1\otimes\hat\tau_2$ in the pseudospin subspaces that correspond
to the two ends of the impurity line. Hereinafter scalar expressions for impurity lines are implied to be multiplied
by $\holOne\otimes\holOne$.

\subsection*{Integrals in $d=2-\varepsilon$ dimensions}

When evaluating diagrams below we use the following values of momentum integrals
in dimension $2-\varepsilon$:

\begin{subequations}
\begin{align}
	\int_\bp\frac{1}{p^2+\Omega^2} &=\frac{C_{2-\varepsilon}}{\varepsilon}\Omega^{-\varepsilon}+\cO(\varepsilon),
	\label{Ia}
	\\
	\int_\bp\frac{1}{(p^2+\Omega^2)^2} &=\frac{1}{2}C_{2-\varepsilon}\Omega^{-2-\varepsilon}+\cO(\varepsilon),
	\label{Ib}
	\\
	\int_\bp\frac{1}{(p^2+\Omega^2)^3} &=\frac{1}{4}C_{2-\varepsilon}\Omega^{-4-\varepsilon}+\cO(\varepsilon),
	\label{Ibb}
	\\
	\int_\bp\frac{1}{(i\Omega-\hbsigma\bp)^2} &=\left(\frac{1}{\varepsilon}-1\right)C_{2-\varepsilon}\Omega^{-\varepsilon}
	+\cO(\varepsilon),
	\label{Ic}
	\\
	\int_\bp\frac{1}{(i\Omega-\hbsigma\bp)^3} &=-\frac{iC_{2-\varepsilon}}{2}\Omega^{-1-\varepsilon}+\cO(\varepsilon),
	\label{Id}
\end{align}
\end{subequations}

\begin{subequations}
\begin{align}
	\int_{\bp,\bq}\frac{1}{(\Omega^2+p^2)(\Omega^2+q^2)[\Omega^2+(\bp+\bq)^2]} &=\cO(1),
	\label{IIa}
	\\
	\int_{\bp,\bq}\frac{\bp\bq}{(\Omega^2+p^2)(\Omega^2+q^2)[\Omega^2+(\bp+\bq)^2]}
	\label{IIb}
	&=-\frac{1}{2}\left(\frac{C_{2-\varepsilon}}{\varepsilon}\Omega^{-\varepsilon}\right)^2+\cO(1),
	\\
	\int_{\bp,\bq}\frac{(\bp\hbsigma)(\bq\hbsigma)}{(\Omega^2+p^2)(\Omega^2+q^2)[\Omega^2+(\bp+\bq)^2]}
	&=-\frac{1}{2}\left(\frac{C_{2-\varepsilon}}{\varepsilon}\Omega^{-\varepsilon}\right)^2+\cO(1),
	\label{IIc}
	\\
	\int_{\bp,\bq}\frac{(\bp\hbsigma)(\bq\hbsigma)}{(\Omega^2+p^2)^2[\Omega^2+(\bp+\bq)^2]}
	&=-\left(\frac{C_{2-\varepsilon}}{\varepsilon}\Omega^{-\varepsilon}\right)^2
	\left(1-\frac{\varepsilon}{2}\right)+\cO(1),
	\label{IId}
	\\
	\int_{\bp,\bq}\frac{(\hbsigma\bp)(\hbsigma\bq)}{(\Omega^2+p^2)^2(\Omega^2+q^2)[\Omega^2+(\bp+\bq)^2]}
	&=\cO(1),
	\label{IIe}
	\\
	\int_{\bp,\bq}\frac{\bp\bq}{(\Omega^2+p^2)^2(\Omega^2+q^2)[\Omega^2+(\bp+\bq)^2]}
	&=\cO(1),
	\label{IIf}
\end{align}
\end{subequations}
where the coefficient $C_{2-\varepsilon}=2(4\pi)^{\frac{\varepsilon}{2}-1}/\Gamma\left(1-\frac{\varepsilon}{2}\right)$
is defined after Eq.~\eqref{DimensDisorder}, and $\int_\bp\ldots...=\int d\bp/(2\pi)^d\ldots$.

Detailed calculations of integrals \eqref{Ia}-\eqref{Ibb} are presented, e.g., in Ref.~\onlinecite{PeskinSchroeder}.
Integrals \eqref{Ic} and \eqref{Id} can be reduced to similar integrals using
$(i\Omega-\hbsigma\bp)^{-1}=-(i\Omega+\hbsigma\bp)/(\Omega^2+p^2)$.

Integral \eqref{IIa} can be evaluated by introducing two Feynman parametrisations\cite{PeskinSchroeder}:
\begin{align}
	\int_{\bp,\bq}\frac{1}{(\Omega^2+p^2)(\Omega^2+q^2)[\Omega^2+(\bp+\bq)^2]}
	=\int_\bq\frac{1}{\Omega^2+q^2}\int_0^1du\int_\bp\frac{1}{[\Omega^2+(1-u)p^2+u(\bp+\bq)^2]^2}
	\nonumber\\
	=\int_\bq\frac{1}{\Omega^2+q^2}\int_0^1du
	\frac{C_d\Gamma\left(2-\frac{d}{2}\right)\Gamma\left(\frac{d}{2}\right)}{2\Gamma(2)}
	[\Omega^2+u(1-u)q^2]^{\frac{d}{2}-2}
	\overset{\varepsilon\ll1}{\approx}
	\frac{C_{2-\varepsilon}}{2}
	\int_0^1 du\int_\bq\frac{1}{(\Omega^2+q^2)\left[\Omega^2+q^2u(1-u)\right]^{2-\frac{d}{2}}}
	\nonumber\\
	\approx\left(\frac{C_{2-\varepsilon}}{2}\right)^2\iint\limits_0^1du\,dt
	\int_\bq \frac{t^{1-\frac{d}{2}}}{\left[\Omega^2+q^2tu(1-u)+(1-t)q^2\right]^{3-\frac{d}{2}}}
	{\approx}
	\left(\frac{C_{2-\varepsilon}}{2}\right)^2\Omega^{2-2\varepsilon}\iint\limits_0^1 dt\,du
	\frac{t^{1-\frac{d}{2}}}{[tu(1-u)+1-t]^\frac{d}{2}}
	=\cO(1)
\end{align}

Integral \eqref{IIb} can be reduced to the previous
integrals by using that $\bp\bq=\frac{1}{2}(\bp+\bq)^2-\frac{1}{2}p^2-\frac{1}{2}q^2$.

In order to evaluate integrals \eqref{IIc}-\eqref{IIf} we note that they are invariant under the interchange
of $\bp$ and $\bq$. They can thus be reduced to the previous integrals by replacing 
$(\bp\hbsigma)(\bq\hbsigma)\rightarrow\frac{1}{2}[(\bp\hbsigma)(\bq\hbsigma)+(\bq\hbsigma)(\bp\hbsigma)]=
\frac{1}{2}(\bp+\bq)^2-\frac{1}{2}p^2-\frac{1}{2}q^2$ or
$\bp\bq\rightarrow\frac{1}{2}(\bp+\bq)^2-\frac{1}{2}p^2-\frac{1}{2}q^2$. 

\section{One-loop renormalisations}

\begin{figure}[htbp]
	\centering
	\includegraphics[width=0.55\textwidth]{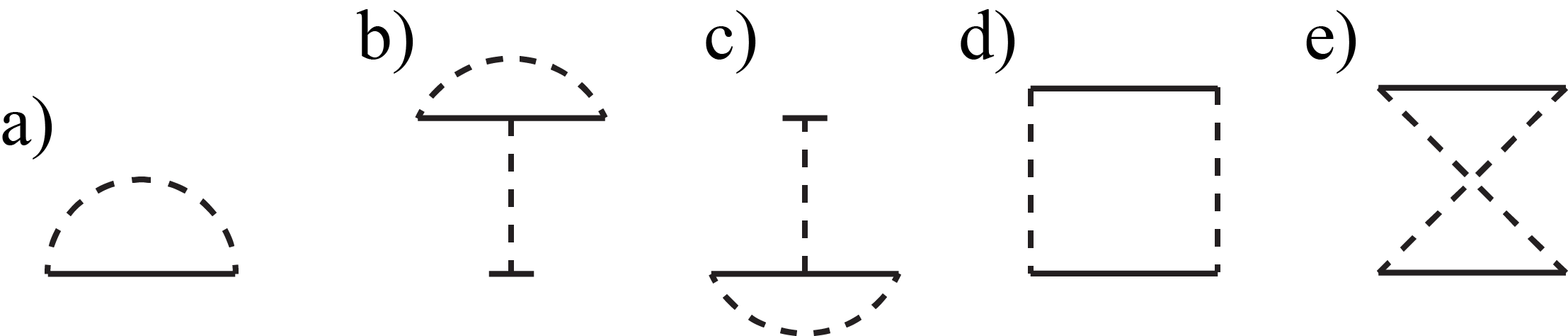}
	\caption{Diagrams for the one-loop renormalisation:
	a) self-energy correction, b-e) vertex corrections.}
	\label{OneLoop}
\end{figure}

One-loop renormalisation is mimicked by the diagrams in Fig.~\ref{OneLoop}.
In what follows expressions in square brackets is our convention for the values of the respective
diagrams.

Diagram \ref{OneLoop}a, the leading-order-in-$\varkappa$ self-energy
of the particles,
is independent of the incoming and outgoing momenta and can be evaluated as
\begin{align}
	[\ref{OneLoop}\text{a}]=\varkappa\int_\bp(i\Omega-\hbsigma\bp)^{-1}
	\overset{\eqref{Ia}}{=}-i\Omega\varkappa\frac{C_{2-\varepsilon}}{\varepsilon}+\cO(\varepsilon).
	\label{OmegaOneLoop}
\end{align}

Diagrams \ref{OneLoop}b-\ref{OneLoop}e mimic the corrections to the disorder strength $\varkappa$.
Because we study the long-wavelength behaviour of the system (at finite frequency), these
diagrams can be evaluated for zero incoming and outgoing momenta, integrating with respect to
the intermediate momenta:
\begin{align}
	[\ref{OneLoop}\text{b}]=[\ref{OneLoop}\text{c}]=\varkappa^2\int_\bp(i\Omega-\hbsigma\bp)^{-2}
	\overset{\eqref{Ic}}{=}\varkappa^2\frac{C_{2-\varepsilon}}{\varepsilon}\Omega^{-\varepsilon}+\cO(1),
	\label{KappaOneLoop1}
\end{align}

\begin{align}
	[\ref{OneLoop}\text{d}]+[\ref{OneLoop}\text{e}]=
	\varkappa^2\int_\bp \frac{1}{i\Omega-\hbsigma\bp}\otimes \frac{1}{i\Omega-\hbsigma\bp}
	+\varkappa^2\int_\bp \frac{1}{i\Omega-\hbsigma\bp}\otimes \frac{1}{i\Omega+\hbsigma\bp}
	=-\varkappa^2\int_\bp\frac{2\Omega^2}{(\Omega^2+p^2)^2}=\cO(1).
	\label{KappaOneLoop2}
\end{align}

To cancel the divergent in $1/\varepsilon$ corrections to the scale $\Omega$
and to the disorder strength $\varkappa$, given by Eqs.~(\ref{OmegaOneLoop}) and (\ref{KappaOneLoop1})-(\ref{KappaOneLoop2})
respectively, we add to the Lagrangian~(\ref{LE}) the counterterm-Lagrangian
\begin{align}
	\cL_{\text{counter}} &=\int \delta^{(1)}\Omega\:\psi^\dagger\psi\: d\br
	+\frac{1}{2}\delta^{(1)}\varkappa\int(\psi^\dagger\psi)^2d\br,
	\\
	\delta^{(1)}\varkappa &=-2\varkappa^2
	\, \frac{C_{2-\varepsilon}}{\varepsilon}\,\Omega^{-\varepsilon},
	\label{KappaCounterOneLoop}
	\\
	\delta^{(1)}\Omega &=-\Omega\varkappa\, \frac{C_{2-\varepsilon}}{\varepsilon}\,\Omega^{-\varepsilon}.
	\label{OmegaCounterOneLoop}
\end{align}

Eqs.~\eqref{KappaCounterOneLoop} and \eqref{OmegaCounterOneLoop} describe the one-loop renormalisation
of the system parameters. We note, that there is no one-loop renormalisation of the particle velocity
(the coefficient before $\hbsigma\bk$ in the Lagrangian).

\begin{figure}[htbp]
	\centering
	\includegraphics[width=0.3\textwidth]{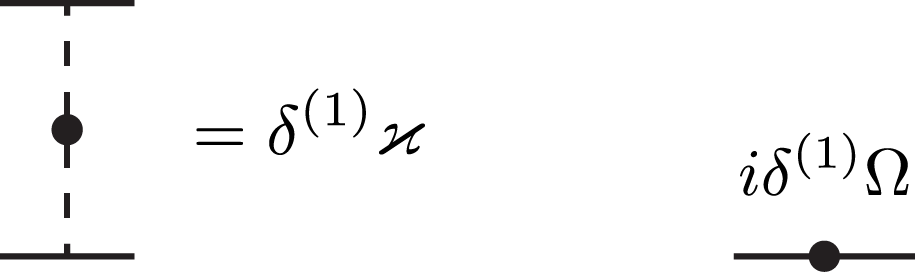}
	\caption{Counterterms coming from the one-loop renormalisation.}
	\label{CounterElem}
\end{figure}

In order to take into account these one-loop corrections when performing the two-loop renormalisation
we introduce additional elements to the diagrammatic technique, Fig.~\ref{CounterElem}.

\section{Two-loop self-energy renormalisation}

The two-loop contribution to the particle self-energy is given by the diagrams in Figs.~\ref{SelfEnergy}
and \ref{SelfEnergyCounter}.
Diagram \ref{SelfEnergy}a depends on the external momentum $\bk$, while the other diagrams are momentum-independent.
The momentum dependency of the two-loop self-energy leads to the renormalisation of the particle
velocity, in addition to the energy scale $\Omega$. 

\subsection{Frequency renormalisation}

\begin{figure}[htbp]
	\centering
	\includegraphics[width=0.54\textwidth]{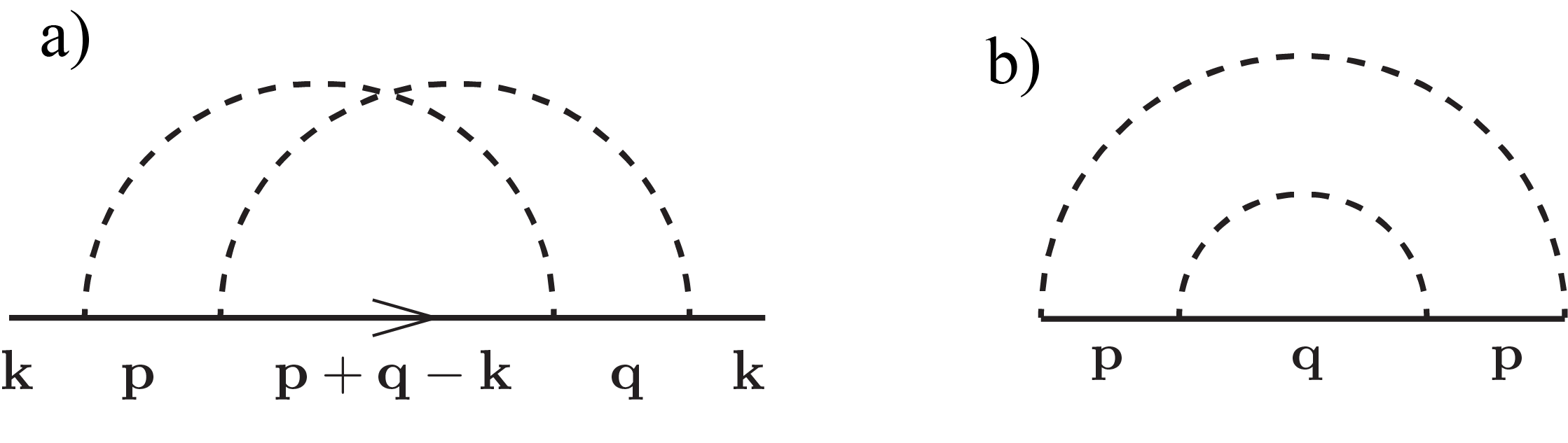}
	\caption{Contributions to the renormalisation of the self-energy.}
	\label{SelfEnergy}
\end{figure}

\begin{figure}[htbp]
	\centering
	\includegraphics[width=0.35\textwidth]{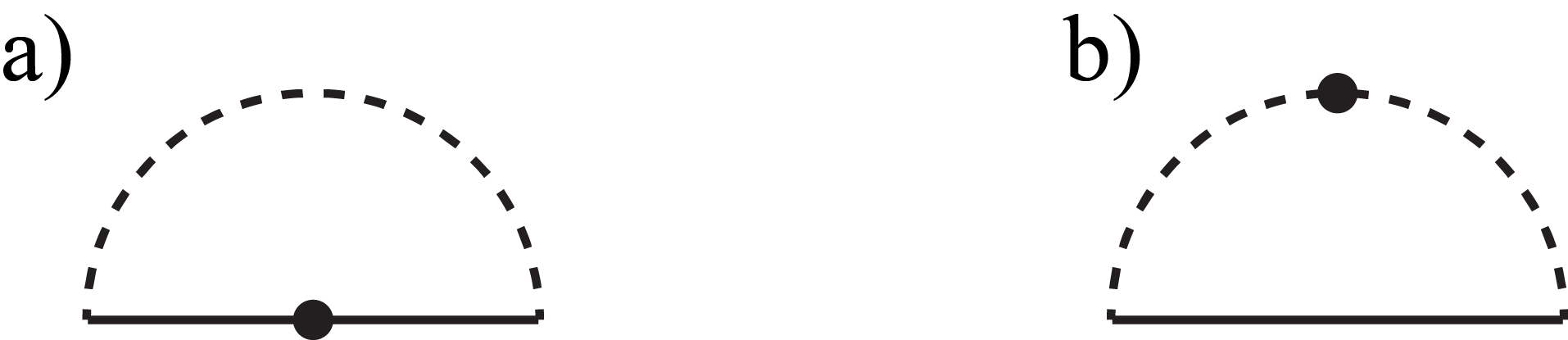}
	\caption{Two-loop contributions to the self-energy that come from one-loop counterterms, Fig.~\ref{CounterElem}.}
	\label{SelfEnergyCounter}
\end{figure}

In order to obtain the two-loop corrections to $\Omega$ it is sufficient to evaluate the diagrams in
Figs.~\ref{SelfEnergy} and \ref{SelfEnergyCounter} for zero external momenta.


Diagrams \ref{SelfEnergy}a and \ref{SelfEnergy}b for $\bk=0$ are given by (in units $\varkappa^2$) 
\begin{align}
	[\ref{SelfEnergy}\text{a}]=
	-\int_{\bp,\bq}\frac{(i\Omega+\hbsigma\bp)\left[i\Omega+\hbsigma(\bp+\bq)\right](i\Omega+\hbsigma\bq)}
	{(\Omega^2+p^2)[\Omega^2+(\bp+\bq)^2](\Omega^2+q^2)}
	\nonumber\\
	\overset{\eqref{IIa}}{=}-i\Omega\int_{\bp,\bq}\frac{(\bp\hbsigma)(\bq\hbsigma)+(\bp\hbsigma)(\bp\hbsigma+\bq\hbsigma)
	+(\bp\hbsigma+\bq\hbsigma)(\bq\hbsigma)}
	{(\Omega^2+p^2)(\Omega^2+q^2)[\Omega^2+(\bp+\bq)^2]}
	+\cO(1) 
	\overset{\eqref{IIc}}{=}
	-\frac{i\Omega}{2}\left(\frac{C_{2-\varepsilon}}{\varepsilon}\Omega^{-\varepsilon}\right)^2+\cO(1)
	\\
	[\ref{SelfEnergy}\text{b}]\overset{\eqref{Ia},\eqref{Ic}}
	=-i\Omega\left(\frac{C_{2-\varepsilon}}{\varepsilon}\Omega^{-\varepsilon}\right)^2(1-\varepsilon)+\cO(1)
\end{align}

Diagrams of the second order in the disorder strength $\varkappa$ that contain the one-loop
counterterms, Fig.~\ref{CounterElem}, are shown in Fig.~\ref{SelfEnergyCounter}.

\begin{align}
	[\ref{SelfEnergyCounter}\text{a}]=\varkappa\:\delta^{(1)}\varkappa\int_\bp(i\Omega-\hbsigma\bp)^{-2}
	\overset{\eqref{Ic},\eqref{KappaCounterOneLoop}}{=}
	i\Omega\varkappa^2\left(\frac{C_{2-\varepsilon}}{\varepsilon}\Omega^{-\varepsilon}\right)^2(1-\varepsilon)+\cO(1)
	\\
	[\ref{SelfEnergyCounter}\text{b}]=2i\Omega\varkappa^2
	\left(\frac{C_{2-\varepsilon}}{\varepsilon}\Omega^{-\varepsilon}\right)^2+\cO(1).
\end{align}

\subsection{Velocity renormalisation}

The velocity renormalisation is determined by the linear-in-$\bk$ contribution to the self-energy.
In order to obtain it, we expand to the linear order the $\bk$-dependent propagator in
diagram \ref{SelfEnergy}a:
\begin{align}
	G_0(i\Omega,\bp+\bq-\bk)\approx G_0(i\Omega,\bp+\bq)-G_0(i\Omega,\bp+\bq)(\hbsigma\bk)G_0(i\Omega,\bp+\bq).
\end{align}

\begin{figure}[htbp]
	\centering
	\includegraphics[width=0.27\textwidth]{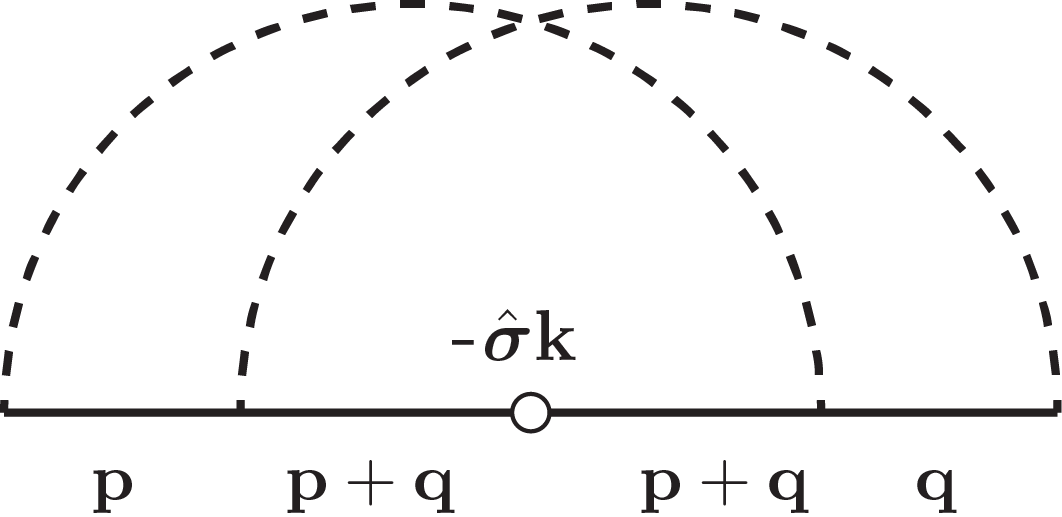}
	\caption{Diagram for the velocity renormalisation.}
	\label{VelocityRenorm}
\end{figure}

The diagram, corresponding to the linear-in-$\bk$ contribution, is shown in Fig.~\ref{VelocityRenorm}.
In units $\varkappa^2$
\begin{align}
	[\ref{VelocityRenorm}]=-\int_{\bp,\bq}
	\frac{1}{i\Omega-\hbsigma\bp}\frac{1}{i\Omega-\hbsigma(\bp+\bq)}\hbsigma\bk
	\frac{1}{i\Omega-\hbsigma(\bp+\bq)}\frac{1}{i\Omega-\hbsigma\bq}
	\nonumber\\
	=-\int_{\bp,\bq}\frac{(i\Omega+\hbsigma\bp)[i\Omega+\hbsigma(\bp+\bq)](\hbsigma\bk)[i\Omega+\hbsigma(\bp+\bq)](i\Omega+\hbsigma\bq)}
	{(\Omega^2+p^2)[\Omega^2+(\bp+\bq)^2]^2(\Omega^2+q^2)}
	\label{VRenormC1}
\end{align}
Using that $(\hbsigma\bp_1)(\hbsigma\bp_2)+(\hbsigma\bp_2)(\hbsigma\bp_1)\equiv 2\bp_1\bp_2$, Eq.~\ref{VRenormC1}
gives
\begin{align}
	[\ref{VelocityRenorm}]=\int_{\bp,\bq}\frac{(i\Omega+\hbsigma\bp)(\hbsigma\bk)(i\Omega+\hbsigma\bq)}
	{(\Omega^2+p^2)[\Omega^2+(\bp+\bq)^2](\Omega^2+q^2)}+I_1
	\label{VRenormC15}
\end{align}	
where
\begin{align}
	I_1=-2\int_{\bp,\bq}\frac{(i\Omega+\hbsigma\bp)[i\Omega+\hbsigma(\bp+\bq)](i\Omega+\hbsigma\bq)}
	{(\Omega^2+p^2)[\Omega^2+(\bp+\bq)^2]^2(\Omega^2+q^2)}
	(\bp+\bq)\bk
	\label{VRenormC2}
\end{align}
Let us demonstrate that $I_1$, Eq.~(\ref{VRenormC2}), does not contain $1/\varepsilon$
or $1/\varepsilon^2$ singularities. Replacing in the numerator
$(i\Omega+\hbsigma\bp)[i\Omega+\hbsigma(\bp+\bq)](i\Omega+\hbsigma\bq)
\rightarrow p^2(\hbsigma\bq)+q^2(\hbsigma\bp)-2\Omega^2\hbsigma(\bp+\bq)$ (as the other contributions vanish)
and using \eqref{IIa} gives
\begin{align}
	I_1=-2\int_{\bp,\bq}
	\frac{p^2(\hbsigma\bq)+q^2(\hbsigma\bp)}{(\Omega^2+p^2)[\Omega^2+(\bp+\bq)^2]^2(\Omega^2+q^2)}
	(\bp+\bq)\bk+\cO(1)
	\label{VrenormC3}
\end{align}
Using that the integral in \eqref{VrenormC3} is invariant with respect to interchanging $\bp$ and $\bq$
and to changing signs of momentum components ($p_\alpha,q_\alpha\rightarrow-p_\alpha,-q_\alpha$),
we replace $(\bq\hbsigma)\cdot[(\bp+\bq)\bk]\equiv\sum_{\alpha,\beta}\sigma_\alpha q_\alpha
(p+q)_\beta k_\beta\rightarrow\sum_\alpha \sigma_\alpha q_\alpha (p+q)_\alpha k_\alpha
\rightarrow\frac{1}{d}\bq(\bp+\bq)\cdot(\hbsigma\bk)$ and arrive at
\begin{align}
	I_1=-\frac{4}{d}\int_{\bp,\bq}
	\frac{p^2 \bq(\bp+\bq)}{(\Omega^2+p^2)[\Omega^2+(\bp+\bq)^2]^2(\Omega^2+q^2)}\hbsigma\bk
	\nonumber\\
	=-\frac{4}{d}\int_{\bp,\bq}\frac{\bq(\bp+\bq)}{(\Omega^2+q^2)[\Omega^2+(\bp+\bq)^2]}
	+\frac{4}{d}\Omega^2\int_{\bp,\bq}\frac{\bq(\bq+\bp)}{(\Omega^2+p^2)[\Omega^2+(\bp+\bq)^2]^2(\Omega^2+q^2)}
	\overset{\eqref{IIf}}{=}\cO(1).
	\label{VRenormC4}
\end{align}

Eqs.~\eqref{VRenormC15}, \eqref{VRenormC2}, and \eqref{VRenormC4} give
\begin{align}
	[\ref{VelocityRenorm}]=\int_{\bp,\bq}\frac{(\hbsigma\bp)(\hbsigma\bk)(\hbsigma\bq)}
	{(\Omega^2+p^2)(\Omega^2+q^2)[\Omega^2+(\bp+\bq)^2]}+\cO(1)
	\nonumber\\
	=\left(\frac{2}{d}-1\right)\int_{\bp,\bq}\frac{\bp\bq}{(\Omega^2+p^2)(\Omega^2+q^2)[\Omega^2+(\bp+\bq)^2]}\hbsigma\bk+\cO(1)
	\overset{\eqref{IIb}}{=}-\frac{1}{4\varepsilon}(C_{2-\varepsilon}\Omega^{-\varepsilon})^2\hbsigma\bk+\cO(1).
\end{align}

\section{Two-loop vertex renormalisation}

\begin{figure}[htbp]
	\centering
	\includegraphics[width=0.9\textwidth]{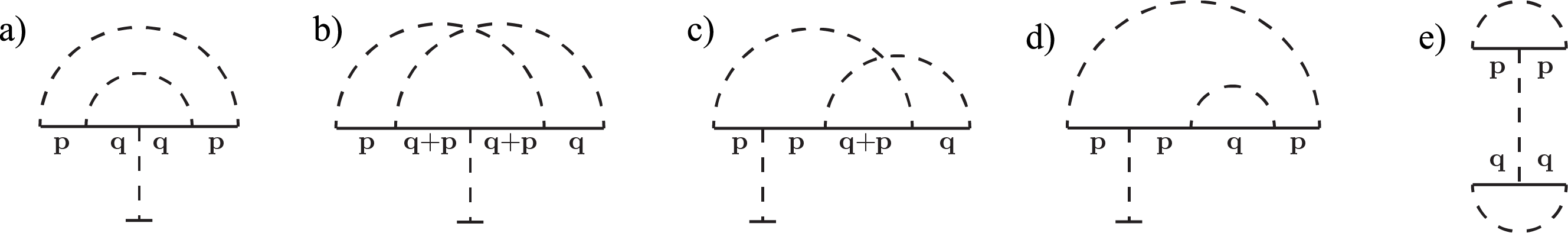}
	\caption{Contributions to the two-loop renormalisation of the impurity line.}
	\label{VertexFive}
\end{figure}

\begin{figure}[htbp]
	\centering
	\includegraphics[width=0.5\textwidth]{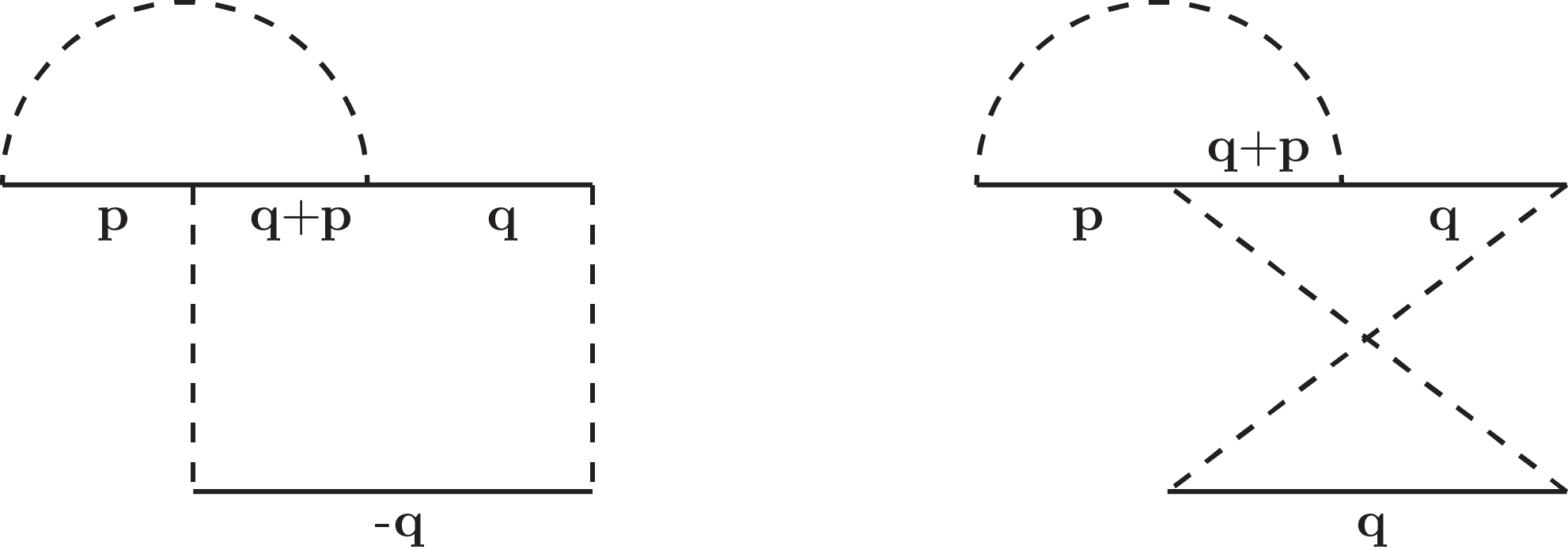}
	\caption{Contribution to the two-loop renormalisation of the impurity line.}
	\label{VertexK}
\end{figure}

\begin{figure}[htbp]
	\centering
	\includegraphics[width=0.4\textwidth]{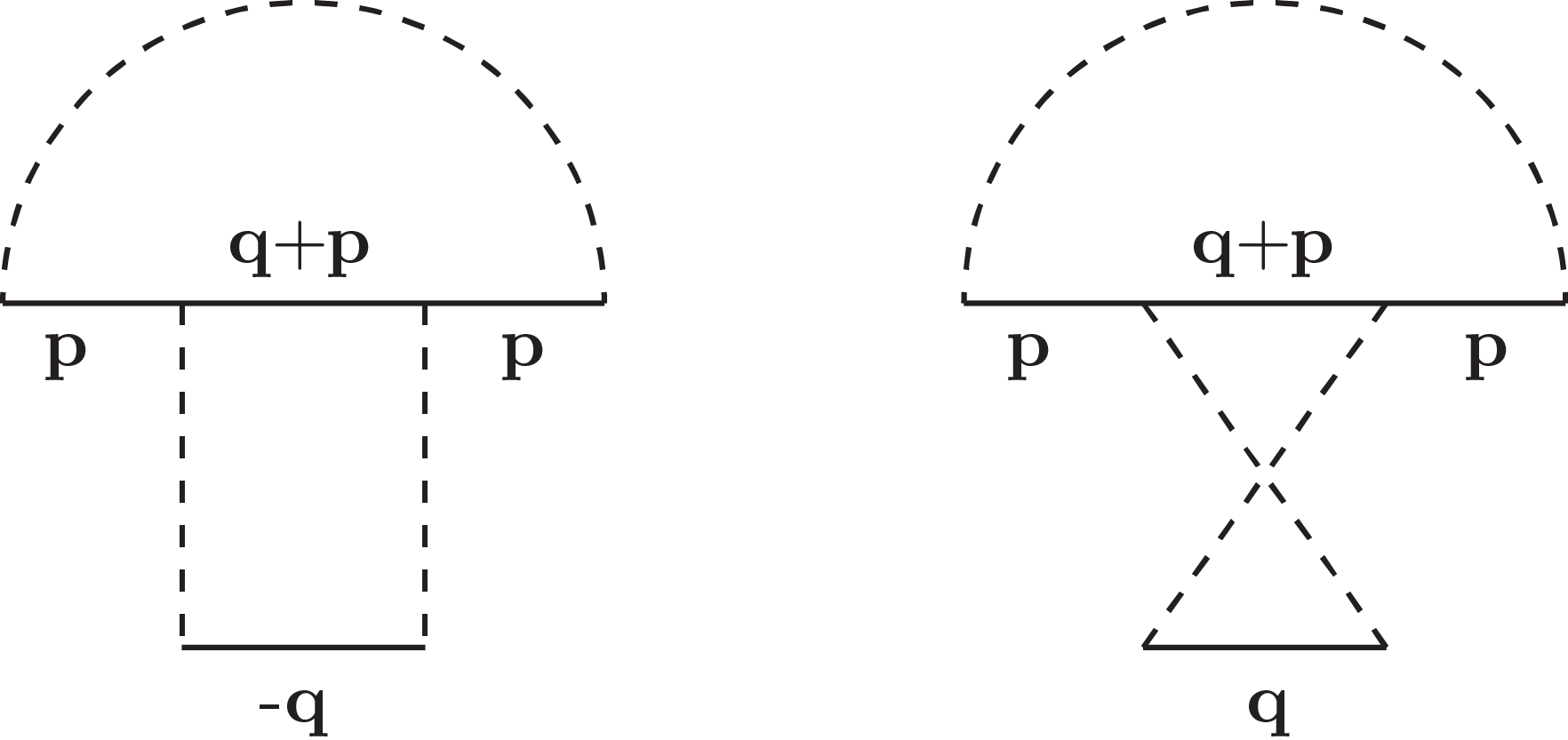}
	\caption{Contribution to the two-loop renormalisation of the impurity line.}
	\label{VertexL}
\end{figure}

\begin{figure}[htbp]
	\centering
	\includegraphics[width=0.4\textwidth]{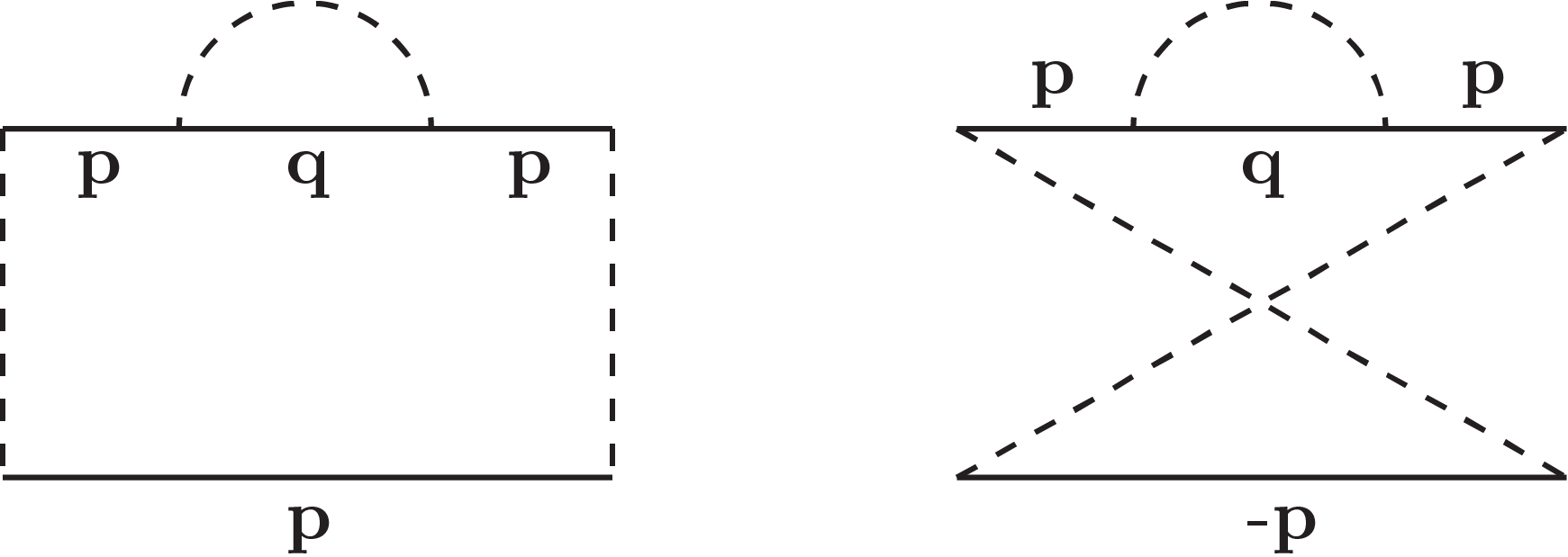}
	\caption{Contribution to the two-loop renormalisation of the impurity line.}
	\label{VertexM}
\end{figure}

\begin{figure}[htbp]
	\centering
	\includegraphics[width=0.4\textwidth]{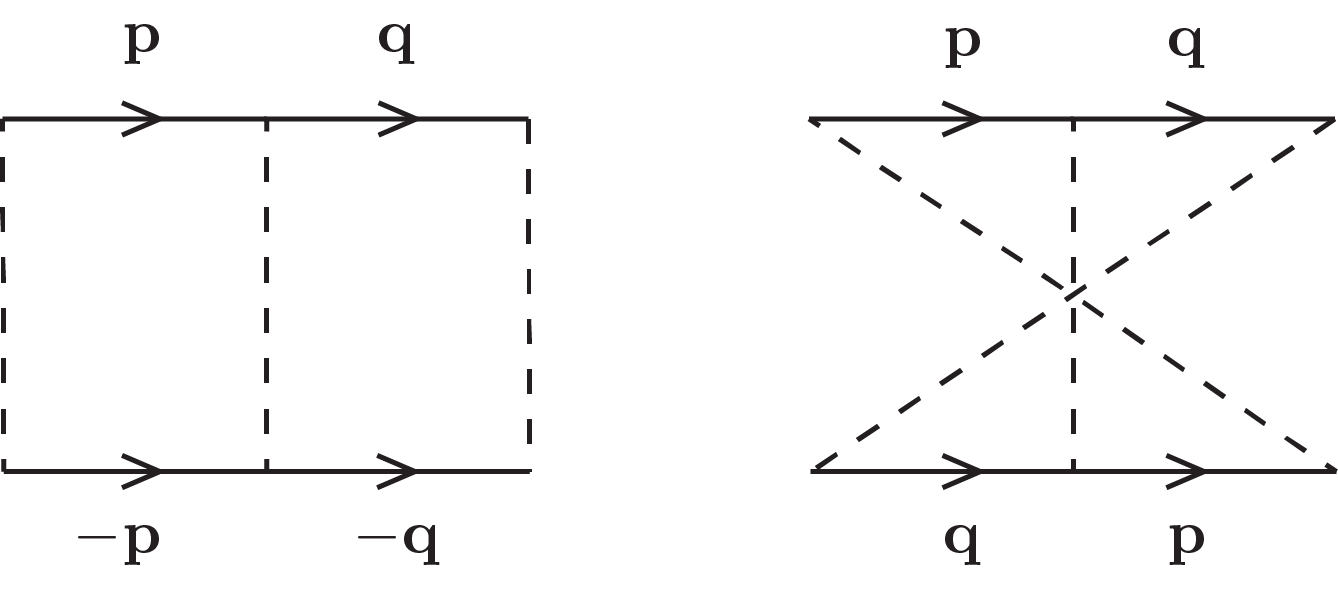}
	\caption{Contribution to the two-loop renormalisation of the impurity line.}
	\label{VertexN}
\end{figure}

\begin{figure}[htbp]
	\centering
	\includegraphics[width=0.4\textwidth]{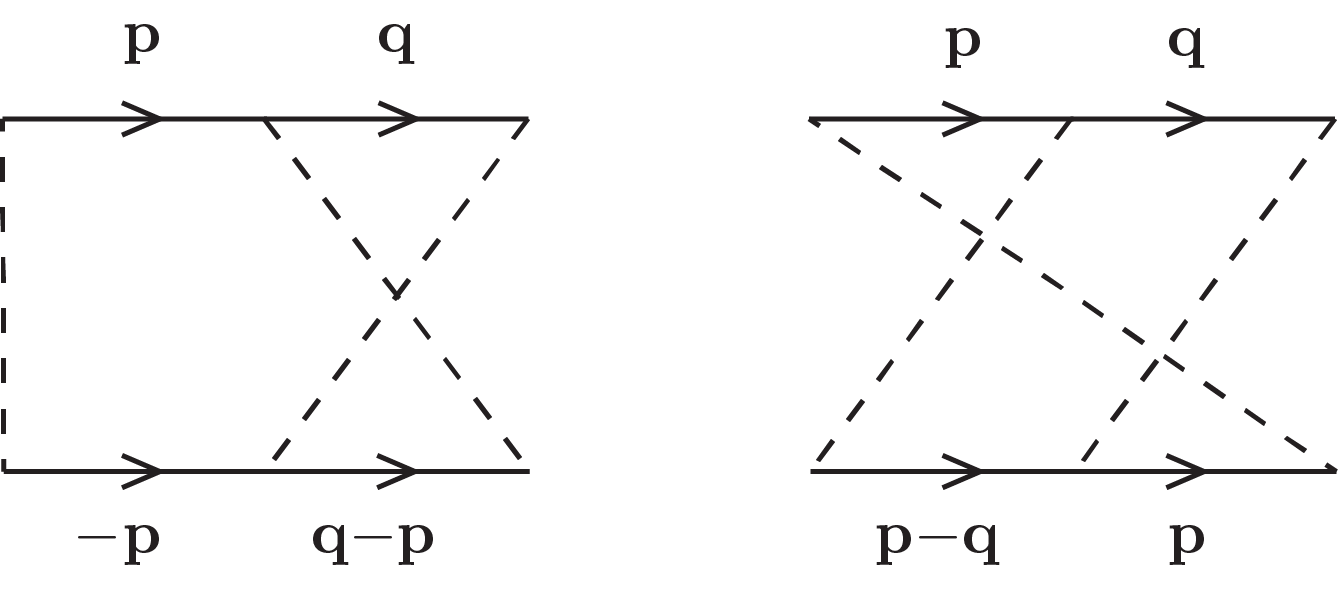}
	\caption{Contribution to the two-loop renormalisation of the impurity line.}
	\label{VertexO}
\end{figure}

\begin{figure}[htbp]
	\centering
	\includegraphics[width=0.5\textwidth]{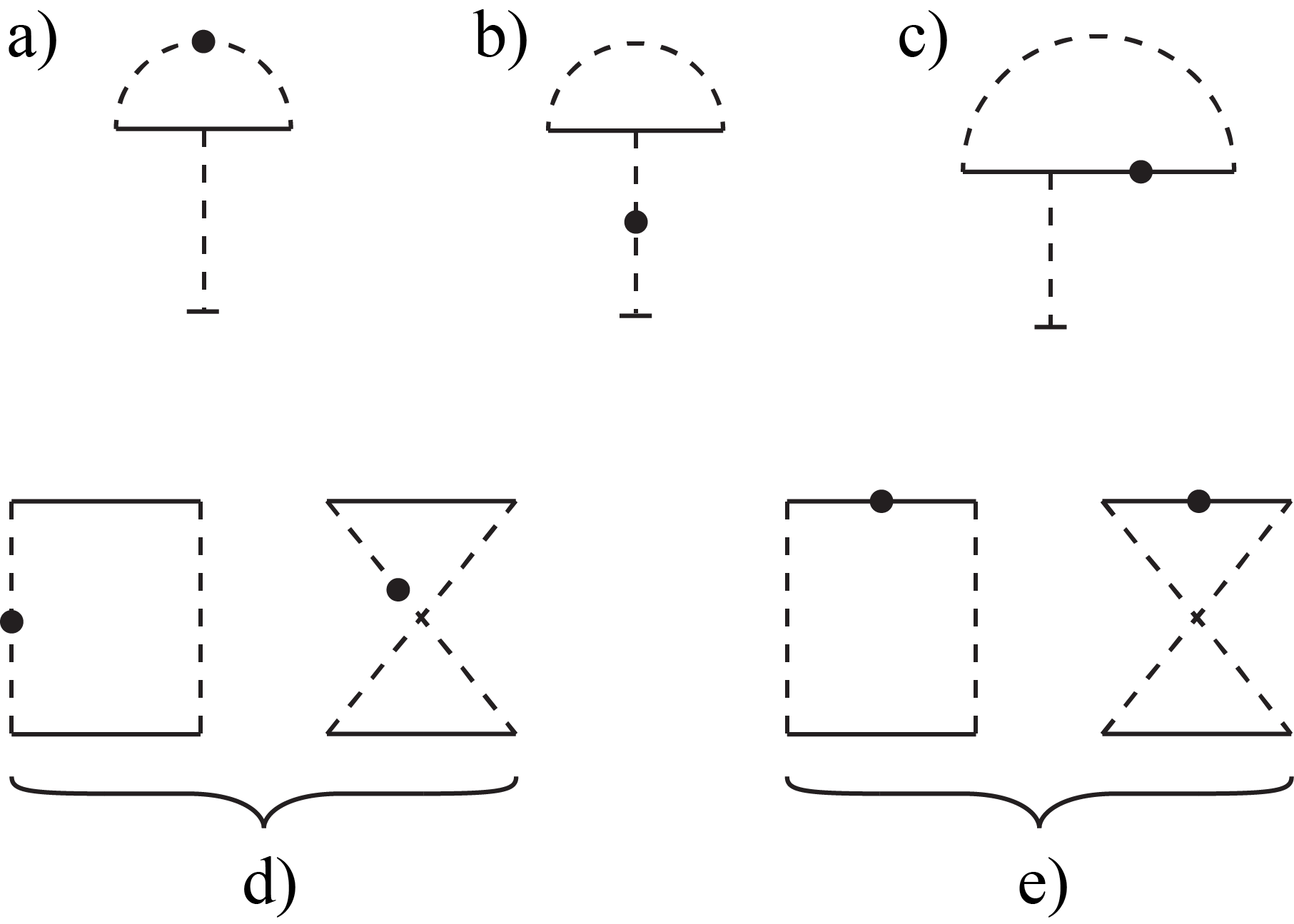}
	\caption{Diagrams for the two-loop renormalisation with one-loop counterterms.}
	\label{TwoLoopCounter}
\end{figure}

The two-loop renormalisation of the disorder strength $\varkappa$ corresponds to the diagrams in
Figs.~\ref{VertexFive}--\ref{TwoLoopCounter} (we show only topologically inequivalent
diagrams). In what immediately follows we present a detailed
calculation of each of these diagrams. For simplicity all expressions for the diagrams are
given in units $\varkappa^3$.

\begin{align}
	[\ref{VertexFive}\text{a}]=\left[\int_\bp\frac{1}{(i\Omega-\hbsigma\bp)^2}\right]^2
	\overset{\eqref{Ia}}{=}\left(C_{2-\varepsilon}\Omega^{-\varepsilon}\right)
	\left(\frac{1}{\varepsilon^2}	-\frac{2}{\varepsilon}\right)+\cO(1)
\end{align}

\begin{align}
	[\ref{VertexFive}\text{b}]=
	\int_{\bp,\bq}\frac{1}{i\Omega-\hbsigma\bp}\frac{1}{[i\Omega-\hbsigma(\bp+\bq)]^2}\frac{1}{i\Omega-\hbsigma\bq}
	=\int_{\bp,\bq}\frac{(i\Omega+\hbsigma\bp)[i\Omega+\hbsigma(\bp+\bq)]^2(i\Omega+\hbsigma\bq)}
	{(\Omega^2+p^2)[\Omega^2+(\bp+\bq)^2]^2(\Omega^2+q^2)}
	\nonumber\\
	=\int_{\bp,\bq}\frac{(\hbsigma\bp)(\hbsigma\bq)(\bp+\bq)^2}{(\Omega^2+p^2)[\Omega^2+(\bp+\bq)^2]^2(\Omega^2+q^2)}
	-\Omega^2\int_{\bp,\bq}
	\frac{2(\hbsigma\bp)(\hbsigma\bp+\hbsigma\bq)+2(\hbsigma\bp+\hbsigma\bq)(\hbsigma\bq)+(\hbsigma\bp)(\hbsigma\bq)}
	{(\Omega^2+p^2)[\Omega^2+(\bp+\bq)^2]^2(\Omega^2+q^2)}+\cO(1)
	\nonumber\\
	\overset{\eqref{IIa}}{=}
	\int_{\bp,\bq}\frac{(\hbsigma\bp)(\hbsigma\bq)}{(\Omega^2+p^2)[\Omega^2+(\bp+\bq)^2](\Omega^2+q^2)}
	-2\Omega^2\int_{\bp,\bq}\frac{(\hbsigma\bp)(\hbsigma\bq)}{(\Omega^2+p^2)[\Omega^2+(\bp+\bq)^2]^2(\Omega^2+q^2)}
	+\cO(1)
	\nonumber\\
	=\int_{\bp,\bq}\frac{(\hbsigma\bp)(\hbsigma\bq)}{(\Omega^2+p^2)[\Omega^2+(\bp+\bq)^2](\Omega^2+q^2)}
	-\Omega^2\int_{\bp,\bq}\frac{(\bp+\bq)^2-p^2-q^2}
	{(\Omega^2+p^2)[\Omega^2+(\bp+\bq)^2]^2(\Omega^2+q^2)}+\cO(1)
	\nonumber\\
	\overset{\eqref{IIc},\eqref{IIb},\eqref{Ib},\eqref{Ic}}{=}
	-\frac{1}{2}\varkappa^3
	(C_{2-\varepsilon}\Omega^{-\varepsilon})^2
	\left(\frac{1}{\varepsilon^2}-\frac{2}{\varepsilon}\right)+\cO(1),
\end{align}

\begin{align}
	[\ref{VertexFive}c]=
	\int_{\bp,\bq}\frac{1}{(i\Omega-\hbsigma\bp)^2}\frac{1}{i\Omega-\hbsigma(\bp+\bq)}\frac{1}{i\Omega-\hbsigma\bq}
	\nonumber\\
	\overset{\bp+\bq\rightarrow\bp,\bq\rightarrow-\bq}{=}
	\int_{\bp,\bq}\frac{[i\Omega+\hbsigma(\bp+\bq)]^2}{[\Omega^2+(\bp+\bq)^2]}
	\frac{i\Omega+\hbsigma\bp}{\Omega^2+p^2}\frac{i\Omega-\hbsigma\bq}{\Omega^2+q^2}
	\nonumber\\
	=-\int_{\bp,\bq}\frac{(\bp+\bq)^2(\hbsigma\bp)(\hbsigma\bq)}
	{(\Omega^2+p^2)[\Omega^2+(\bp+\bq)^2]^2(\Omega^2+q^2)}
	\nonumber\\
	-\Omega^2
	\int_{\bp,\bq}\frac{2(\hbsigma\bp+\hbsigma\bq)(\hbsigma\bp)+2(\hbsigma\bp+\hbsigma\bq)(\hbsigma\bq)-(\hbsigma\bp)(\hbsigma\bq)}
	{(\Omega^2+p^2)[\Omega^2+(\bp+\bq)^2]^2(\Omega^2+q^2)}+\cO(1)
	\nonumber\\
	=-[\ref{VertexFive}b]+\cO(1)=
	\frac{1}{2}\varkappa^3
	(C_{2-\varepsilon}\Omega^{-\varepsilon})^2
	\left(\frac{1}{\varepsilon^2}-\frac{2}{\varepsilon}\right)+\cO(1)
\end{align}

\begin{align}
	[\ref{VertexFive}\text{d}]=\int_\bp\left(\frac{1}{i\Omega-\hbsigma\bp}\right)^3
	\int_\bq\frac{1}{i\Omega-\hbsigma\bp}\overset{\eqref{Ia},\eqref{Id}}{=}
	=-\frac{1}{2\varepsilon}\,(C_{2-\varepsilon}\Omega^{-\varepsilon})^2+\cO(1)
\end{align}

\begin{align}
	[\ref{VertexFive}\text{e}]
	=\int_\bp\frac{1}{(i\Omega-\hbsigma\bp)^2}\otimes\int_\bq\frac{1}{(i\Omega-\hbsigma\bq)^2}
	\overset{\eqref{Ic}}{=}\left(C_{2-\varepsilon}\Omega^{-\varepsilon}\right)^2
	\left(\frac{1}{\varepsilon^2}-\frac{2}{\varepsilon}\right)+\cO(1)
\end{align}

\begin{align}
	[\ref{VertexK}]=
	\int_{\bp,\bq}\frac{1}{i\Omega-\hbsigma\bp}\frac{1}{i\Omega-\hbsigma(\bp+\bq)}\frac{1}{i\Omega-\hbsigma\bq}
	\otimes\left(\frac{1}{i\Omega-\hbsigma\bq}+\frac{1}{i\Omega+\hbsigma\bq}\right)
	\nonumber\\
	=2i\Omega\int
	\frac{(i\Omega+\hbsigma\bp)[i\Omega+\hbsigma(\bp+\bq)](i\Omega+\hbsigma\bq)}
	{(\Omega^2+p^2)[\Omega^2+(\bp+\bq)^2](\Omega^2+q^2)^2}
	=-2\Omega^2\int_{\bp,\bq}
	\frac{(\hbsigma\bp)(\hbsigma\bq)+(\bp+\bq)^2}{(\Omega^2+p^2)[\Omega^2+(\bp+\bq)^2](\Omega^2+q^2)^2}
	+\cO(1)
	\nonumber\\
	=-\Omega^2\int_{\bp,\bq}
	\frac{3(\bp+\bq)^2-p^2-q^2}{(\Omega^2+p^2)[\Omega^2+(\bp+\bq)^2](\Omega^2+q^2)^2}
	\nonumber\\
	=-3\Omega^2\int_{\bp,\bq}\frac{1}{(\Omega^2+p^2)(\Omega^2+q^2)^2}
	+\Omega^2\int_{\bp,\bq}\frac{1}{[\Omega^2+(\bq+\bp)^2](\Omega^2+q^2)^2}+\cO(1)
	\nonumber\\
	\overset{\eqref{Ia},\eqref{Ib}}{=}-\frac{1}{\varepsilon}
	\left(C_{2-\varepsilon}\Omega^{-\varepsilon}\right)^2+\cO(1)
\end{align}

\begin{align}
	[\ref{VertexL}]=
	\int_{\bp,\bq}\frac{1}{i\Omega-\hbsigma\bp}\frac{1}{i\Omega-\hbsigma(\bp+\bq)}\frac{1}{i\Omega-\hbsigma\bp}
	\otimes\left(\frac{1}{i\Omega-\hbsigma\bq}+\frac{1}{i\Omega+\hbsigma\bq}\right)
	\nonumber\\
	=2i\Omega\int
	\frac{(i\Omega+\hbsigma\bp)[i\Omega+\hbsigma(\bp+\bq)](i\Omega+\hbsigma\bp)}
	{(\Omega^2+p^2)^2[\Omega^2+(\bp+\bq)^2](\Omega^2+q^2)}
	=-2\Omega^2\int_{\bp,\bq}
	\frac{(\bp^2+\bq^2)-p^2-q^2}{(\Omega^2+p^2)^2[\Omega^2+(\bp+\bq)^2](\Omega^2+q^2)}
	+\cO(1)
	\nonumber\\
	\overset{\eqref{IIa}}{=}
	-2\Omega^2\int_{\bp,\bq}\frac{1}{(\Omega^2+p^2)^2(\Omega^2+q^2)}
	+2\Omega^2\int_{\bp,\bq}\frac{1}{(\Omega^2+p^2)^2[\Omega^2+(\bp+\bq)^2]}+\cO(1)
	=\cO(1)
\end{align}

\begin{align}
	[\ref{VertexM}]=
	\int_{\bp,\bq}\frac{1}{i\Omega-\hbsigma\bp}\frac{1}{i\Omega-\hbsigma\bq}\frac{1}{i\Omega-\hbsigma\bp}
	\otimes
	\left(\frac{1}{i\Omega-\hbsigma\bp}+\frac{1}{i\Omega+\hbsigma\bp}\right)
	\nonumber\\
	\overset{\eqref{Ia}}{=}
	\left[-i\Omega\frac{C_{2-\varepsilon}}{\varepsilon}\Omega^{-\varepsilon}+\cO(\varepsilon)\right]\cdot
	2i\Omega\int_\bp\frac{(2i\Omega+\hbsigma\bp)^2}{(\Omega^2+p^2)^3}
	=2\Omega^2\left[\frac{C_{2-\varepsilon}}{\varepsilon}+\cO(\varepsilon)\right]\cdot
	\left[\int_\bp\frac{1}{(\Omega^2+p^2)^2}-2\Omega^2\int_\bp\frac{1}{(\Omega^2+p^2)^3}\right]
	\nonumber\\
	\overset{\eqref{Ib},\eqref{Ib}}{=}
	\frac{1}{\varepsilon}\cdot\cO(\varepsilon)=\cO(1)
\end{align}

The sum of the two diagrams in Fig.~\ref{VertexN} is given by
\begin{align}
	[\ref{VertexN}]=\int_{\bp,\bq}\frac{1}{i\Omega-\hbsigma\bp}\frac{1}{i\Omega-\hbsigma\bq}
	\otimes\frac{1}{i\Omega+\hbsigma\bp}\frac{1}{i\Omega+\hbsigma\bq}
	+\int_{\bp,\bq}\frac{1}{i\Omega-\hbsigma\bp}\frac{1}{i\Omega-\hbsigma\bq}
	\otimes\frac{1}{i\Omega-\hbsigma\bq}\frac{1}{i\Omega-\hbsigma\bp}
	\nonumber\\
	=\int_{\bp,\bq}\frac{(i\Omega+\hbsigma\bp)(i\Omega+\hbsigma\bq)\otimes
	\left[-2\Omega^2+(\hbsigma\bp)(\hbsigma\bq)+(\hbsigma\bq)(\hbsigma\bp)\right]}
	{(\Omega^2+p^2)^2(\Omega^2+q^2)^2}
	=2\int_{\bp,\bq}\frac{[-\Omega^2+(\hbsigma\bp)(\hbsigma\bq)]\otimes
	(-\Omega^2+\bp\bq)}
	{(\Omega^2+p^2)^2(\Omega^2+q^2)^2}
	\nonumber\\
	=2\int_{\bp,\bq}\frac{(\hbsigma\bp)(\hbsigma\bq)\otimes\bp\bq}{(\Omega^2+p^2)^2(\Omega^2+q^2)^2}+\cO(1)
	\label{VertNIntegralInterm}
\end{align}
Using that in the tensor product $(\hbsigma\bp)(\hbsigma\bq)\otimes\bp\bq=
\sum_{\alpha,\beta,\gamma}\sigma_\alpha p_\alpha\sigma_\beta q_\beta\otimes p_\gamma q_\gamma$ 
only the terms with $\alpha=\beta=\gamma$ contribute
to the integral (\ref{VertNIntegralInterm}) we replace 
$(\hbsigma\bp)(\hbsigma\bq)\otimes\bp\bq\rightarrow\sum_\alpha p_\alpha^2 q_\alpha^2\rightarrow \frac{1}{d}p^2 q^2$:
\begin{align}
	[\ref{VertexN}]=\frac{2}{d}\left[\int_\bp\frac{p^2}{(\Omega^2+p^2)^2}\right]^2
	=\frac{2}{d}\left[\int_\bp\frac{1}{\Omega^2+p^2}-\Omega^2\int_\bp\frac{1}{(\Omega^2+p^2)^2}\right]^2
	\nonumber\\
	=\frac{2}{2-\varepsilon}
	\left[\frac{1}{\varepsilon}C_{2-\varepsilon}\Omega^{-\varepsilon}
	-\frac{1}{2}C_{2-\varepsilon}\Omega^{-\varepsilon}+\cO(\varepsilon)\right]^2
	=\left(\frac{1}{\varepsilon^2}-\frac{1}{2\varepsilon}\right)
	\left(C_{2-\varepsilon}\Omega^{-\varepsilon}\right)^2+\cO(1).
\end{align}

\begin{align}
	[\ref{VertexO}]=\int_{\bp,\bq}\frac{1}{i\Omega-\hbsigma\bp}\frac{1}{i\Omega-\hbsigma\bq}\otimes
	\frac{1}{i\Omega+\hbsigma\bp}\frac{1}{i\Omega+\hbsigma(\bp-\bq)}+
	\int_{\bp,\bq}\frac{1}{i\Omega-\hbsigma\bp}\frac{1}{i\Omega-\hbsigma\bq}\otimes
	\frac{1}{i\Omega-\hbsigma(\bp-\bq)}\frac{1}{i\Omega-\hbsigma\bp}
	\nonumber\\
	=\int_{\bp,\bq}
	\frac{(i\Omega+\hbsigma\bp)(i\Omega+\hbsigma\bq)\otimes[-2\Omega^2+(\hbsigma\bp)(\hbsigma\bp-\hbsigma\bq)+(\hbsigma\bp-\hbsigma\bq)(\hbsigma\bp)]}
	{(\Omega^2+p^2)^2(\Omega^2+q^2)[\Omega^2+(\bp-\bq)^2]}
	\nonumber\\
	=2\int_{\bp,\bq}
	\frac{(i\Omega+\hbsigma\bp)(i\Omega+\hbsigma\bq)\otimes[-\Omega^2+\bp(\bp-\bq)]}
	{(\Omega^2+p^2)^2(\Omega^2+q^2)[\Omega^2+(\bp-\bq)^2]}
	=2\int_{\bp,\bq}\frac{[-\Omega^2+(\hbsigma\bp)(\hbsigma\bq)][-\Omega^2+\bp(\bp-\bq)]}
	{(\Omega^2+p^2)^2(\Omega^2+q^2)[\Omega^2+(\bp-\bq)^2]}
	\nonumber\\
	=2\int_{\bp,\bq}\frac{(\hbsigma\bp)(\hbsigma\bq)\otimes\bp(\bp-\bq)}{(\Omega^2+p^2)^2(\Omega^2+q^2)[\Omega^2+(\bp-\bq)^2]}
	-2\Omega^2\int_{\bp,\bq}\frac{(\hbsigma\bp)(\hbsigma\bq)\otimes\holOne}{(\Omega^2+p^2)^2(\Omega^2+q^2)[\Omega^2+(\bp-\bq)^2]}
	\nonumber\\
	-2\Omega^2\int_{\bp,\bq}\frac{\holOne\otimes\bp(\bp-\bq)}{(\Omega^2+p^2)^2(\Omega^2+q^2)[\Omega^2+(\bp-\bq)^2]}
	+\cO(1)
	\overset{\eqref{IIe},\eqref{IIf}}{=}
	\int_{\bp,\bq}\frac{(\hbsigma\bp)(\hbsigma\bq)\otimes[p^2+(\bp-\bq)^2-q^2]}
	{(\Omega^2+p^2)^2(\Omega^2+q^2)[\Omega^2+(\bp-\bq)^2]}+\cO(1)
	\nonumber\\
	=\int_{\bp,\bq}\frac{(\hbsigma\bp)(\hbsigma\bq)\otimes\holOne}{(\Omega^2+p^2)(\Omega^2+q^2)[\Omega^2+(\bp-\bq)^2]}
	-\int_{\bp,\bq}\frac{(\hbsigma\bp)(\hbsigma\bq)\otimes\holOne}{(\Omega^2+p^2)^2[\Omega^2+(\bp-\bq)^2]}
	\nonumber\\
	-\Omega^2\int_{\bp,\bq}
	\frac{(\hbsigma\bp)(\hbsigma\bq)}{(\Omega^2+p^2)^2(\Omega^2+q^2)[\Omega^2+(\bp-\bq)^2]}+\cO(1)
	\nonumber\\
	\overset{\eqref{IIc},\eqref{IId}}{=}
	\frac{1}{2}\left(\frac{C_{2-\varepsilon}}{\varepsilon}\Omega^{-\varepsilon}\right)^2
	-\left(\frac{C_{2-\varepsilon}}{\varepsilon}\Omega^{-\varepsilon}\right)^2
	\left(1-\frac{\varepsilon}{2}\right)+\cO(1)
	=-\frac{1}{2}(C_{2-\varepsilon}\Omega^{-\varepsilon})^2
	\left(\frac{1}{\varepsilon^2}+\frac{1}{\varepsilon}\right)+\cO(1)
\end{align}

Diagrams for the two-loop renormalisation that contain the one-loop counterterms, Fig.~\ref{CounterElem},
are shown in Fig.~\ref{TwoLoopCounter}. Diagram \ref{TwoLoopCounter}c
can be evaluated straightforwardly using Eq.~(\ref{Id}),
and the other diagrams-- similarly to the corresponding one-loop diagrams in Fig.~\ref{OneLoop}:
\begin{align}
	[\ref{TwoLoopCounter}a]=[\ref{TwoLoopCounter}b]=-2\varkappa^3 (C_{2-\varepsilon}\Omega^{-\varepsilon})^2
	\left(\frac{1}{\varepsilon^2}-\frac{1}{\varepsilon}\right)+\cO(1),
\end{align}

\begin{align}
	[\ref{TwoLoopCounter}c]=\frac{1}{2}\varkappa^3(C_{2-\varepsilon}\Omega^{-\varepsilon})^2/\varepsilon+\cO(1),
\end{align}

\begin{align}
	[\ref{TwoLoopCounter}d]=2\varkappa^3(C_{2-\varepsilon}\Omega^{-\varepsilon})^2/\varepsilon+\cO(1),
\end{align}

\begin{align}
	[\ref{TwoLoopCounter}e]=\cO(1).
\end{align}


\section{RG equations}

In the previous sections we calculated in two loops perturbative corrections to observable couplings
$\Omega$, $\varkappa$, and the quasiparticle velocity (coefficient before $\hbsigma\hat\bk$) that come from
the action \eqref{LE} and contain divergent contributions $\propto 1/\varepsilon$ and $1/\varepsilon^2$.
In order to cancel these divergencies, following the minimal subtraction scheme\cite{PeskinSchroeder},
we add to the action \eqref{LE} the counterterm Lagrangian
\begin{align}
	\cL_{\text{counter}} &=-i\int \psi^\dagger	\left[\delta(i\Omega)-\delta(\hbsigma\hat\bk)\right]\psi\: d\br
	+\frac{1}{2}\delta\varkappa\int\left(\psi^\dagger\psi\right)^2d\br,
\end{align}
where
\begin{align}
	\delta(i\Omega)=[\ref{OneLoop}a]+[\ref{SelfEnergy}a]_{\bk=0}+[\ref{SelfEnergy}b]
	+[\ref{SelfEnergyCounter}a]+[\ref{SelfEnergyCounter}b]
	=i\Omega\left[-\frac{1}{\varepsilon}\varkappa C_{2-\varepsilon}\Omega^{-\varepsilon}
	+\frac{3}{2}\frac{1}{\varepsilon^2}
	\left(\varkappa C_{2-\varepsilon}\Omega^{-\varepsilon}\right)^2
	\right],
	\label{OmegaDivergent}
\end{align}
\begin{align}
	\delta(\hbsigma\bk)=-[\ref{VelocityRenorm}]
	=\frac{1}{4\varepsilon}(\varkappa C_{2-\varepsilon}\Omega^{-\varepsilon})^2\hbsigma\bk
\end{align}
\begin{align}
	\delta\varkappa=-[\ref{OneLoop}b-e]-2\times[\ref{VertexFive}a]-2\times[\ref{VertexFive}b]
	-4\times[\ref{VertexFive}c]-4\times[\ref{VertexFive}d]-[\ref{VertexFive}e]-4\times[\ref{VertexK}]
	-2\times[\ref{VertexL}]-2\times[\ref{VertexM}]-[\ref{VertexN}]-2\times[\ref{VertexO}]
	\nonumber\\
	-2\times[\ref{TwoLoopCounter}a]-2\times[\ref{TwoLoopCounter}b]
	-4\times[\ref{TwoLoopCounter}c]-2\times[\ref{TwoLoopCounter}d]-2\times[\ref{TwoLoopCounter}e]
	=\varkappa\left[-\frac{2}{\varepsilon}\varkappa C_{2-\varepsilon}\Omega^{-\varepsilon}
	-\left(-\frac{4}{\varepsilon^2}+\frac{1}{2\varepsilon}\right)\left(\varkappa C_{2-\varepsilon}\Omega^{-\varepsilon}\right)^2\right]
	\label{KappaDivergent}
\end{align}
where we kept only the terms divergent at $\varepsilon\rightarrow0$ and took into account the numbers
of topologically equivalent diagrams.

Introducing
dimensionless disorder strength, Eq.~(\ref{DimensDisorder}), the full Lagrangian (\ref{Action1})
of the system can be rewritten in terms of the observable variables $\varkappa$ and $\Omega$ as
\begin{align}
	\cL=-i\int\psi^\dagger
	\left[i\Omega\left(1-\frac{\gamma}{2\varepsilon}
	+\frac{3}{8}\frac{\gamma^2}{\varepsilon^2}\right)\right.
	\left.-\hbsigma\hat\bk
	\left(1+\frac{\gamma^2}{16\varepsilon}\right)
	\right]\psi \,d\br
	-\frac{\gamma\Omega^\varepsilon}{4C_{2-\varepsilon}}
	\left(1-\frac{\gamma}{\varepsilon}-\frac{\gamma^2}{8\varepsilon}
	+\frac{\gamma^2}{\varepsilon^2}\right)\int(\psi^\dagger\psi)^2d\br.
	\label{LagrangianFinal}
\end{align}

The renormalised observables $\psi$, $\gamma$, and $\Omega$ can be related to the ``bare''
$\Phi$, $\varkappa_0$ and $\omega$ by comparing the Lagrangian \eqref{LagrangianFinal} with
the ``bare'' one, Eq.~(\ref{Action0}): 
\begin{align}
	Z &= 1+\frac{\gamma^2}{16\varepsilon},\\
	\varkappa_0 &= \frac{\Omega^\varepsilon}{2C_{2-\varepsilon}}\gamma
	\left(1-\frac{\gamma}{\varepsilon}+\frac{\gamma^2}{\varepsilon^2}-\frac{\gamma^2}{4\varepsilon}\right),
	\label{KappaOGamma}
	\\
	\omega &=\Omega\left(1-\frac{\gamma}{2\varepsilon}+\frac{3\gamma^2}{8\varepsilon^2}-\frac{\gamma^2}{16\varepsilon}\right),
	\label{OmegaBare}
\end{align}
where $Z$ describes the rescaling of the particle wavefunctons: $\psi=\Phi/Z^\frac{1}{2}$.

The energy scale $\Omega$
sets the characteristic momentum of long-wavelength behaviour
of the system. In order to obtain the RG flow of the renormalised disorder strength
$\gamma$ as a function of $\Omega$ for a given bare disorder strength $\varkappa_0$, we require 
\begin{align}
	\frac{\partial\varkappa_0}{\partial\ln\Omega}=0,
\end{align}
which, together with Eq.~(\ref{KappaOGamma}), gives the RG equation \eqref{GammaRGE}.

RG equation \eqref{OmegaRGE} follows straightforwardly from Eq.~\eqref{OmegaBare}.


\begin{thebibliography}{39}
\expandafter\ifx\csname natexlab\endcsname\relax\def\natexlab#1{#1}\fi
\expandafter\ifx\csname bibnamefont\endcsname\relax
  \def\bibnamefont#1{#1}\fi
\expandafter\ifx\csname bibfnamefont\endcsname\relax
  \def\bibfnamefont#1{#1}\fi
\expandafter\ifx\csname citenamefont\endcsname\relax
  \def\citenamefont#1{#1}\fi
\expandafter\ifx\csname url\endcsname\relax
  \def\url#1{\texttt{#1}}\fi
\expandafter\ifx\csname urlprefix\endcsname\relax\def\urlprefix{URL }\fi
\providecommand{\bibinfo}[2]{#2}
\providecommand{\eprint}[2][]{\url{#2}}

\bibitem[{\citenamefont{Fradkin}(1986{\natexlab{a}})}]{Fradkin1}
\bibinfo{author}{\bibfnamefont{E.}~\bibnamefont{Fradkin}},
  \bibinfo{journal}{Phys. Rev. B} \textbf{\bibinfo{volume}{33}},
  \bibinfo{pages}{3263} (\bibinfo{year}{1986}{\natexlab{a}}).

\bibitem[{\citenamefont{Fradkin}(1986{\natexlab{b}})}]{Fradkin2}
\bibinfo{author}{\bibfnamefont{E.}~\bibnamefont{Fradkin}},
  \bibinfo{journal}{Phys. Rev. B} \textbf{\bibinfo{volume}{33}},
  \bibinfo{pages}{3257} (\bibinfo{year}{1986}{\natexlab{b}}).

\bibitem[{\citenamefont{Kobayashi et~al.}(2014)\citenamefont{Kobayashi,
  Ohtsuki, Imura, and Herbut}}]{Herbut}
\bibinfo{author}{\bibfnamefont{K.}~\bibnamefont{Kobayashi}},
  \bibinfo{author}{\bibfnamefont{T.}~\bibnamefont{Ohtsuki}},
  \bibinfo{author}{\bibfnamefont{K.-I.} \bibnamefont{Imura}}, \bibnamefont{and}
  \bibinfo{author}{\bibfnamefont{I.~F.} \bibnamefont{Herbut}},
  \bibinfo{journal}{Phys. Rev. Lett.} \textbf{\bibinfo{volume}{112}},
  \bibinfo{pages}{016402} (\bibinfo{year}{2014}).

\bibitem[{\citenamefont{Syzranov
  et~al.}(2015{\natexlab{a}})\citenamefont{Syzranov, Radzihovsky, and
  Gurarie}}]{Syzranov:Weyl}
\bibinfo{author}{\bibfnamefont{S.~V.} \bibnamefont{Syzranov}},
  \bibinfo{author}{\bibfnamefont{L.}~\bibnamefont{Radzihovsky}},
  \bibnamefont{and} \bibinfo{author}{\bibfnamefont{V.}~\bibnamefont{Gurarie}},
  \bibinfo{journal}{Phys. Rev. Lett.} \textbf{\bibinfo{volume}{114}},
  \bibinfo{pages}{166601} (\bibinfo{year}{2015}{\natexlab{a}}).

\bibitem[{\citenamefont{Syzranov
  et~al.}(2015{\natexlab{b}})\citenamefont{Syzranov, Gurarie, and
  Radzihovsky}}]{Syzranov:unconv}
\bibinfo{author}{\bibfnamefont{S.~V.} \bibnamefont{Syzranov}},
  \bibinfo{author}{\bibfnamefont{V.}~\bibnamefont{Gurarie}}, \bibnamefont{and}
  \bibinfo{author}{\bibfnamefont{L.}~\bibnamefont{Radzihovsky}},
  \bibinfo{journal}{Phys. Rev. B} \textbf{\bibinfo{volume}{91}},
  \bibinfo{pages}{035133} (\bibinfo{year}{2015}{\natexlab{b}}).

\bibitem[{\citenamefont{G{\"a}rttner et~al.}(2015)\citenamefont{G{\"a}rttner,
  Syzranov, Rey, Gurarie, and Radzihovsky}}]{Garttner:longrange}
\bibinfo{author}{\bibfnamefont{M.}~\bibnamefont{G{\"a}rttner}},
  \bibinfo{author}{\bibfnamefont{S.~V.} \bibnamefont{Syzranov}},
  \bibinfo{author}{\bibfnamefont{A.~M.} \bibnamefont{Rey}},
  \bibinfo{author}{\bibfnamefont{V.}~\bibnamefont{Gurarie}}, \bibnamefont{and}
  \bibinfo{author}{\bibfnamefont{L.}~\bibnamefont{Radzihovsky}},
  \bibinfo{journal}{Phys. Rev. B} \textbf{\bibinfo{volume}{92}},
  \bibinfo{pages}{041406(R)} (\bibinfo{year}{2015}).

\bibitem[{\citenamefont{Huang et~al.}(2015)\citenamefont{Huang, Xu, Belopolski,
  Lee, Chang, Wang, Alidoust, Bian, Neupane, Zhang et~al.}}]{ZHasan:TaAs}
\bibinfo{author}{\bibfnamefont{S.-M.} \bibnamefont{Huang}},
  \bibinfo{author}{\bibfnamefont{S.-Y.} \bibnamefont{Xu}},
  \bibinfo{author}{\bibfnamefont{I.}~\bibnamefont{Belopolski}},
  \bibinfo{author}{\bibfnamefont{C.-C.} \bibnamefont{Lee}},
  \bibinfo{author}{\bibfnamefont{G.}~\bibnamefont{Chang}},
  \bibinfo{author}{\bibfnamefont{B.}~\bibnamefont{Wang}},
  \bibinfo{author}{\bibfnamefont{N.}~\bibnamefont{Alidoust}},
  \bibinfo{author}{\bibfnamefont{G.}~\bibnamefont{Bian}},
  \bibinfo{author}{\bibfnamefont{M.}~\bibnamefont{Neupane}},
  \bibinfo{author}{\bibfnamefont{C.}~\bibnamefont{Zhang}},
  \bibnamefont{et~al.}, \bibinfo{journal}{Nature Comm.}
  \textbf{\bibinfo{volume}{6}}, \bibinfo{pages}{7373} (\bibinfo{year}{2015}).

\bibitem[{\citenamefont{Xu et~al.}(2015)\citenamefont{Xu, Belopolski, Alidoust,
  Neupane, Bian, Zhang, Sankar, Chang, Yuan, Lee et~al.}}]{ZHasan:TaAs2}
\bibinfo{author}{\bibfnamefont{S.-Y.} \bibnamefont{Xu}},
  \bibinfo{author}{\bibfnamefont{I.}~\bibnamefont{Belopolski}},
  \bibinfo{author}{\bibfnamefont{N.}~\bibnamefont{Alidoust}},
  \bibinfo{author}{\bibfnamefont{M.}~\bibnamefont{Neupane}},
  \bibinfo{author}{\bibfnamefont{G.}~\bibnamefont{Bian}},
  \bibinfo{author}{\bibfnamefont{C.}~\bibnamefont{Zhang}},
  \bibinfo{author}{\bibfnamefont{R.}~\bibnamefont{Sankar}},
  \bibinfo{author}{\bibfnamefont{G.}~\bibnamefont{Chang}},
  \bibinfo{author}{\bibfnamefont{Z.}~\bibnamefont{Yuan}},
  \bibinfo{author}{\bibfnamefont{C.-C.} \bibnamefont{Lee}},
  \bibnamefont{et~al.}, \bibinfo{journal}{Science}
  \textbf{\bibinfo{volume}{349}}, \bibinfo{pages}{613} (\bibinfo{year}{2015}).

\bibitem[{\citenamefont{Lv et~al.}(2015)\citenamefont{Lv, Weng, Fu, Wang, Miao,
  Ma, Richard, Huang, Zhao, Chen et~al.}}]{Weng:PhotCrystWSM}
\bibinfo{author}{\bibfnamefont{B.~Q.} \bibnamefont{Lv}},
  \bibinfo{author}{\bibfnamefont{H.~M.} \bibnamefont{Weng}},
  \bibinfo{author}{\bibfnamefont{B.~B.} \bibnamefont{Fu}},
  \bibinfo{author}{\bibfnamefont{X.~P.} \bibnamefont{Wang}},
  \bibinfo{author}{\bibfnamefont{H.}~\bibnamefont{Miao}},
  \bibinfo{author}{\bibfnamefont{J.}~\bibnamefont{Ma}},
  \bibinfo{author}{\bibfnamefont{P.}~\bibnamefont{Richard}},
  \bibinfo{author}{\bibfnamefont{X.~C.} \bibnamefont{Huang}},
  \bibinfo{author}{\bibfnamefont{L.~X.} \bibnamefont{Zhao}},
  \bibinfo{author}{\bibfnamefont{G.~F.} \bibnamefont{Chen}},
  \bibnamefont{et~al.}, \bibinfo{journal}{Phys. Rev. X}
  \textbf{\bibinfo{volume}{5}}, \bibinfo{pages}{031013} (\bibinfo{year}{2015}).

\bibitem[{\citenamefont{Goswami and Chakravarty}(2011)}]{Goswami:TIRG}
\bibinfo{author}{\bibfnamefont{P.}~\bibnamefont{Goswami}} \bibnamefont{and}
  \bibinfo{author}{\bibfnamefont{S.}~\bibnamefont{Chakravarty}},
  \bibinfo{journal}{Phys. Rev. Lett.} \textbf{\bibinfo{volume}{107}},
  \bibinfo{pages}{196803} (\bibinfo{year}{2011}).

\bibitem[{\citenamefont{Roy and Sarma}(2014)}]{RoyDasSarma}
\bibinfo{author}{\bibfnamefont{B.}~\bibnamefont{Roy}} \bibnamefont{and}
  \bibinfo{author}{\bibfnamefont{S.~D.} \bibnamefont{Sarma}},
  \bibinfo{journal}{Phys. Rev. B} \textbf{\bibinfo{volume}{90}},
  \bibinfo{pages}{241112(R)} (\bibinfo{year}{2014}).

\bibitem[{\citenamefont{Ryi and Nomura}(2012)}]{RyuNomura}
\bibinfo{author}{\bibfnamefont{S.}~\bibnamefont{Ryi}} \bibnamefont{and}
  \bibinfo{author}{\bibfnamefont{K.}~\bibnamefont{Nomura}},
  \bibinfo{journal}{Phys. Rev. B} \textbf{\bibinfo{volume}{85}},
  \bibinfo{pages}{155138} (\bibinfo{year}{2012}).

\bibitem[{\citenamefont{Shindou and Murakami}(2009)}]{ShindouMurakami}
\bibinfo{author}{\bibfnamefont{R.}~\bibnamefont{Shindou}} \bibnamefont{and}
  \bibinfo{author}{\bibfnamefont{S.}~\bibnamefont{Murakami}},
  \bibinfo{journal}{Phys. Rev. B} \textbf{\bibinfo{volume}{79}},
  \bibinfo{pages}{045321} (\bibinfo{year}{2009}).

\bibitem[{\citenamefont{Ominato and Koshino}(2014)}]{OminatoKoshino}
\bibinfo{author}{\bibfnamefont{Y.}~\bibnamefont{Ominato}} \bibnamefont{and}
  \bibinfo{author}{\bibfnamefont{M.}~\bibnamefont{Koshino}},
  \bibinfo{journal}{Phys. Rev. B} \textbf{\bibinfo{volume}{89}},
  \bibinfo{pages}{054202} (\bibinfo{year}{2014}).

\bibitem[{\citenamefont{Aleiner and Efetov}(2006)}]{AleinerEfetov}
\bibinfo{author}{\bibfnamefont{I.~L.} \bibnamefont{Aleiner}} \bibnamefont{and}
  \bibinfo{author}{\bibfnamefont{K.~B.} \bibnamefont{Efetov}},
  \bibinfo{journal}{Phys. Rev. Lett.} \textbf{\bibinfo{volume}{97}},
  \bibinfo{pages}{236801} (\bibinfo{year}{2006}).

\bibitem[{\citenamefont{Ostrovsky et~al.}(2006)\citenamefont{Ostrovsky, Gornyi,
  and Mirlin}}]{OstrovskyGornyMirlin}
\bibinfo{author}{\bibfnamefont{P.~M.} \bibnamefont{Ostrovsky}},
  \bibinfo{author}{\bibfnamefont{I.~V.} \bibnamefont{Gornyi}},
  \bibnamefont{and} \bibinfo{author}{\bibfnamefont{A.~D.}
  \bibnamefont{Mirlin}}, \bibinfo{journal}{Phys. Rev. B}
  \textbf{\bibinfo{volume}{74}}, \bibinfo{pages}{235443}
  (\bibinfo{year}{2006}).

\bibitem[{\citenamefont{Liu et~al.}(2015)\citenamefont{Liu, Ohtsuki, and
  Shindou}}]{LiuOhtsuki:LateNumerics}
\bibinfo{author}{\bibfnamefont{S.}~\bibnamefont{Liu}},
  \bibinfo{author}{\bibfnamefont{T.}~\bibnamefont{Ohtsuki}}, \bibnamefont{and}
  \bibinfo{author}{\bibfnamefont{R.}~\bibnamefont{Shindou}}
  (\bibinfo{year}{2015}), \bibinfo{note}{arXiv:1507.02381}.

\bibitem[{\citenamefont{Bera et~al.}(2016)\citenamefont{Bera, Sau, and
  Roy}}]{Bera:Weyl}
\bibinfo{author}{\bibfnamefont{S.}~\bibnamefont{Bera}},
  \bibinfo{author}{\bibfnamefont{J.~D.} \bibnamefont{Sau}}, \bibnamefont{and}
  \bibinfo{author}{\bibfnamefont{B.}~\bibnamefont{Roy}} (\bibinfo{year}{2016}),
  \bibinfo{note}{arXiv:1507.07551}.

\bibitem[{\citenamefont{Sbierski et~al.}(2015)\citenamefont{Sbierski,
  Bergholtz, and Brouwer}}]{Brouwer:exponents}
\bibinfo{author}{\bibfnamefont{B.}~\bibnamefont{Sbierski}},
  \bibinfo{author}{\bibfnamefont{E.~J.} \bibnamefont{Bergholtz}},
  \bibnamefont{and} \bibinfo{author}{\bibfnamefont{P.~W.}
  \bibnamefont{Brouwer}}, \bibinfo{journal}{Phys. Rev. B}
  \textbf{\bibinfo{volume}{92}}, \bibinfo{pages}{115145}
  (\bibinfo{year}{2015}).

\bibitem[{\citenamefont{Pixley et~al.}(2015)\citenamefont{Pixley, Goswami, and
  Sarma}}]{Pixley:ExactZ}
\bibinfo{author}{\bibfnamefont{J.~H.} \bibnamefont{Pixley}},
  \bibinfo{author}{\bibfnamefont{P.}~\bibnamefont{Goswami}}, \bibnamefont{and}
  \bibinfo{author}{\bibfnamefont{S.~D.} \bibnamefont{Sarma}}
  (\bibinfo{year}{2015}), \bibinfo{note}{arXiv:1505.07938v2}.

\bibitem[{\citenamefont{Schuessler et~al.}(2009)\citenamefont{Schuessler,
  Ostrovsky, Gornyi, and Mirlin}}]{Ostrovsky:grapheneRG}
\bibinfo{author}{\bibfnamefont{A.}~\bibnamefont{Schuessler}},
  \bibinfo{author}{\bibfnamefont{P.~M.} \bibnamefont{Ostrovsky}},
  \bibinfo{author}{\bibfnamefont{I.~V.} \bibnamefont{Gornyi}},
  \bibnamefont{and} \bibinfo{author}{\bibfnamefont{A.~D.}
  \bibnamefont{Mirlin}}, \bibinfo{journal}{Phys. Rev. B}
  \textbf{\bibinfo{volume}{79}}, \bibinfo{pages}{075405}
  (\bibinfo{year}{2009}).

\bibitem[{\citenamefont{Wetzel}(1985)}]{Wetzel:twoloop}
\bibinfo{author}{\bibfnamefont{W.}~\bibnamefont{Wetzel}},
  \bibinfo{journal}{Phys. Lett.} \textbf{\bibinfo{volume}{153B}},
  \bibinfo{pages}{297} (\bibinfo{year}{1985}).

\bibitem[{\citenamefont{Ludwig}(1987)}]{Ludwig:twoloop}
\bibinfo{author}{\bibfnamefont{A.~W.~W.} \bibnamefont{Ludwig}},
  \bibinfo{journal}{Nucl. Phys. B} \textbf{\bibinfo{volume}{285}},
  \bibinfo{pages}{97} (\bibinfo{year}{1987}).

\bibitem[{\citenamefont{Luperini and Rossi}(1991)}]{Rossi1}
\bibinfo{author}{\bibfnamefont{C.}~\bibnamefont{Luperini}} \bibnamefont{and}
  \bibinfo{author}{\bibfnamefont{P.}~\bibnamefont{Rossi}},
  \bibinfo{journal}{Ann. Phys.} \textbf{\bibinfo{volume}{212}},
  \bibinfo{pages}{371} (\bibinfo{year}{1991}).

\bibitem[{\citenamefont{Bondi et~al.}(1990)\citenamefont{Bondi, Curci, Paffuti,
  and Rossi}}]{Rossi2}
\bibinfo{author}{\bibfnamefont{A.}~\bibnamefont{Bondi}},
  \bibinfo{author}{\bibfnamefont{G.}~\bibnamefont{Curci}},
  \bibinfo{author}{\bibfnamefont{G.}~\bibnamefont{Paffuti}}, \bibnamefont{and}
  \bibinfo{author}{\bibfnamefont{P.}~\bibnamefont{Rossi}},
  \bibinfo{journal}{Ann. Phys.} \textbf{\bibinfo{volume}{199}},
  \bibinfo{pages}{268} (\bibinfo{year}{1990}).

\bibitem[{\citenamefont{Tracas and Vlachos}(1990)}]{TracasVlachos}
\bibinfo{author}{\bibfnamefont{N.~D.} \bibnamefont{Tracas}} \bibnamefont{and}
  \bibinfo{author}{\bibfnamefont{N.~D.} \bibnamefont{Vlachos}},
  \bibinfo{journal}{Phys. Lett.} \textbf{\bibinfo{volume}{236}},
  \bibinfo{pages}{333} (\bibinfo{year}{1990}).

\bibitem[{\citenamefont{Ludwig and Wiese}(2003)}]{Ludwig:Thirring}
\bibinfo{author}{\bibfnamefont{A.~W.~W.} \bibnamefont{Ludwig}}
  \bibnamefont{and} \bibinfo{author}{\bibfnamefont{K.~J.} \bibnamefont{Wiese}},
  \bibinfo{journal}{Nucl. Phys. B} \textbf{\bibinfo{volume}{661}},
  \bibinfo{pages}{577} (\bibinfo{year}{2003}).

\bibitem[{\citenamefont{Efetov}(1999)}]{Efetov:book}
\bibinfo{author}{\bibfnamefont{K.~B.} \bibnamefont{Efetov}},
  \emph{\bibinfo{title}{Supersymetry in Disorder and Chaos}}
  (\bibinfo{publisher}{Cambridge University Press}, \bibinfo{address}{New
  York}, \bibinfo{year}{1999}).

\bibitem[{\citenamefont{Kamenev}(2011)}]{Kamenev:book}
\bibinfo{author}{\bibfnamefont{A.}~\bibnamefont{Kamenev}},
  \emph{\bibinfo{title}{Field Theory of Non-Equilibrium Systems}}
  (\bibinfo{publisher}{Cambridge Univ. Press}, \bibinfo{year}{2011}).

\bibitem[{\citenamefont{Belitz and Kirkpatrick}(1994)}]{BelitzKirkpatrick}
\bibinfo{author}{\bibfnamefont{D.}~\bibnamefont{Belitz}} \bibnamefont{and}
  \bibinfo{author}{\bibfnamefont{T.~R.} \bibnamefont{Kirkpatrick}},
  \bibinfo{journal}{Rev. Mod. Phys.} \textbf{\bibinfo{volume}{66}},
  \bibinfo{pages}{261} (\bibinfo{year}{1994}).

\bibitem[{\citenamefont{Nielsen and Ninomiya}(1981)}]{NielsenNinomiya}
\bibinfo{author}{\bibfnamefont{H.~B.} \bibnamefont{Nielsen}} \bibnamefont{and}
  \bibinfo{author}{\bibfnamefont{M.}~\bibnamefont{Ninomiya}},
  \bibinfo{journal}{Nuclear Physics B} \textbf{\bibinfo{volume}{185}},
  \bibinfo{pages}{20} (\bibinfo{year}{1981}).

\bibitem[{\citenamefont{Peskin and Schroeder}(1975)}]{PeskinSchroeder}
\bibinfo{author}{\bibfnamefont{M.~E.} \bibnamefont{Peskin}} \bibnamefont{and}
  \bibinfo{author}{\bibfnamefont{D.~V.} \bibnamefont{Schroeder}},
  \emph{\bibinfo{title}{An Introduction To Quantum Field Theory}}
  (\bibinfo{publisher}{Addison-Wesley}, \bibinfo{year}{1975}).

\bibitem[{\citenamefont{Sbierski et~al.}(2014)\citenamefont{Sbierski, Pohl,
  Bergholtz, and Brouwer}}]{Brouwer:WSMcond}
\bibinfo{author}{\bibfnamefont{B.}~\bibnamefont{Sbierski}},
  \bibinfo{author}{\bibfnamefont{G.}~\bibnamefont{Pohl}},
  \bibinfo{author}{\bibfnamefont{E.~J.} \bibnamefont{Bergholtz}},
  \bibnamefont{and} \bibinfo{author}{\bibfnamefont{P.~W.}
  \bibnamefont{Brouwer}}, \bibinfo{journal}{Phys. Rev. Lett.}
  \textbf{\bibinfo{volume}{113}}, \bibinfo{pages}{026602}
  (\bibinfo{year}{2014}).

\bibitem[{\citenamefont{Shapourian and Hughes}(2016)}]{Shapourian:PhaseDiagr}
\bibinfo{author}{\bibfnamefont{H.}~\bibnamefont{Shapourian}} \bibnamefont{and}
  \bibinfo{author}{\bibfnamefont{T.~L.} \bibnamefont{Hughes}},
  \bibinfo{journal}{Phys. Rev. B} \textbf{\bibinfo{volume}{93}},
  \bibinfo{pages}{075108} (\bibinfo{year}{2016}).

\bibitem[{\citenamefont{Chayes et~al.}(1986)\citenamefont{Chayes, Chayes,
  Fisher, and Spencer}}]{ChayesChayes}
\bibinfo{author}{\bibfnamefont{J.~T.} \bibnamefont{Chayes}},
  \bibinfo{author}{\bibfnamefont{L.}~\bibnamefont{Chayes}},
  \bibinfo{author}{\bibfnamefont{D.~S.} \bibnamefont{Fisher}},
  \bibnamefont{and} \bibinfo{author}{\bibfnamefont{T.}~\bibnamefont{Spencer}},
  \bibinfo{journal}{Phys. Rev. Lett.} \textbf{\bibinfo{volume}{57}},
  \bibinfo{pages}{2999} (\bibinfo{year}{1986}).

\bibitem[{\citenamefont{Harris}(1974)}]{Harris}
\bibinfo{author}{\bibfnamefont{A.~B.} \bibnamefont{Harris}},
  \bibinfo{journal}{J. Phys. C} \textbf{\bibinfo{volume}{7}},
  \bibinfo{pages}{3082} (\bibinfo{year}{1974}).

\bibitem[{\citenamefont{Pixley et~al.}(2016)\citenamefont{Pixley, Goswami, and
  Sarma}}]{Pixley:ExactZpubliahed}
\bibinfo{author}{\bibfnamefont{J.~H.} \bibnamefont{Pixley}},
  \bibinfo{author}{\bibfnamefont{P.}~\bibnamefont{Goswami}}, \bibnamefont{and}
  \bibinfo{author}{\bibfnamefont{S.~D.} \bibnamefont{Sarma}},
  \bibinfo{journal}{Phys. Rev. B} \textbf{\bibinfo{volume}{93}},
  \bibinfo{pages}{085103} (\bibinfo{year}{2016}).

\bibitem[{\citenamefont{Roy and Sarma}(2016)}]{RoyDasSarma:erratum}
\bibinfo{author}{\bibfnamefont{B.}~\bibnamefont{Roy}} \bibnamefont{and}
  \bibinfo{author}{\bibfnamefont{S.~D.} \bibnamefont{Sarma}}
  (\bibinfo{year}{2016}), \bibinfo{note}{arXiv:1602.03470}.

\bibitem[{\citenamefont{Abrikosov et~al.}(1975)\citenamefont{Abrikosov, Gorkov,
  and Dzyaloshinski}}]{AGD}
\bibinfo{author}{\bibfnamefont{A.~A.} \bibnamefont{Abrikosov}},
  \bibinfo{author}{\bibfnamefont{L.~P.} \bibnamefont{Gorkov}},
  \bibnamefont{and} \bibinfo{author}{\bibfnamefont{I.~E.}
  \bibnamefont{Dzyaloshinski}}, \emph{\bibinfo{title}{Methods of Quantum Field
  Theory in Statistical Physics}} (\bibinfo{publisher}{Dover, New York},
  \bibinfo{year}{1975}).

\end{thebibliography}
\end{document}